\documentclass[useAMS,usenatbib]{mn2e}

\usepackage{epsfig}
\usepackage{myaasmacros}
\usepackage{enumerate}
\usepackage{amsmath}
\usepackage{amsfonts}
\usepackage{comment}
\usepackage{color}

\def\lesssim{\lower.5ex\hbox{$\; \buildrel < \over \sim \;$}}
\def\gtrsim{\lower.5ex\hbox{$\; \buildrel > \over \sim \;$}}

\title[Cosmological simulations of black hole growth]
{Cosmological simulations of black hole growth: AGN luminosities and
  downsizing}   
\author[Hirschmann et al.]{Michaela Hirschmann$^{1}$\thanks{E-mail:
mhirsch@oats.inaf.it}, Klaus Dolag$^{2,3}$,  Alexandro Saro$^{2}$,
Lisa Karin Bachmann$^{2}$, 
                               \newauthor  Stefano Borgani$^{1,4,5}$,
                               Andreas Burkert$^{2,6}$ \\ 
$^{1}$INAF - Astronomical Observatory of Trieste, via G.B. Tiepolo 11,
I-34143 Trieste, Italy\\
$^{2}$Universit\"ats Sternwarte M\"unchen, Scheinerstr.1, D-81679
M\"unchen, Germany\\ 
$^{3}$Max-Plank-Institute f\"ur Astrophysik, Karl-Schwarzschild
Strasse 1, D-85740 Garching, Germany\\
$^{4}$Astronomy Unit, Department of Physics, University of Trieste,
via G.B. Tiepolo 11, I-34131 Trieste, Italy\\ 
$^{5}$INFN-National Institute for Nuclear Physics, Via Valerio 2,
I-34127 Trieste, Italy\\ 
$^{6}$Max-Planck-Institut f\"ur extraterrestrische Physik (MPE),
Giessenbachstr. 1, 85748 Garching, Germany} 

\begin{document}

\date{Accepted ???. Received ??? in original form ???}

\pagerange{\pageref{firstpage}--\pageref{lastpage}} \pubyear{2002}

\maketitle

\label{firstpage}
\begin{abstract}
In this study, we present a detailed, statistical analysis of black
hole growth and the evolution of active galactic nuclei (AGN) using
cosmological hydrodynamic simulations run down to $z=0$. The
simulations self-consistently follow radiative cooling, star
formation, metal enrichment, black hole growth and associated
feedback processes from both supernovae typeII/Ia and AGN. We consider
two simulation runs, one with a large co-moving  volume of $(500\
\mathrm{Mpc})^3$ and one with a smaller volume of  $(68\
\mathrm{Mpc})^3$ but with a by a factor of almost 20 higher mass
resolution. We compare the predicted statistical properties of AGN 
with results from large observational surveys. Consistently with
previous results, our simulations can widely match observed black hole
properties of the local Universe. Furthermore, our simulations can
successfully reproduce the evolution of the bolometric AGN luminosity
function for both the low-luminosity and the high-luminosity end up to
$z=3.0$, only at $z=1.5-2.5$, the low luminosity end is over-estimated
by up to 1~dex. 
In addition, the smaller but higher resolution run is able to match
the observational data of the low bolometric luminosity end at higher
redshifts $z=3-4$. We also perform a direct comparison with the
observed soft and hard X-ray luminosity functions of AGN, including an
empirical correction for a torus-level obscuration, and find a
similarly good agreement. These results nicely demonstrate that the
observed ``anti-hierarchical'' trend in the AGN  number density
evolution (i.e. the number densities of luminous AGN peak at higher
redshifts than those of faint AGN) is self-consistently predicted by our
simulations. Implications of this downsizing behaviour on active black
holes, their masses and Eddington-ratios are discussed. Overall, the
downsizing behaviour in the AGN number density as a function of
redshift can be mainly attributed to the evolution of the gas density
in the resolved vicinity of a (massive) black hole (which is
  depleted with evolving time as a consequence of star formation and
  AGN feedback).
\end{abstract} 


\section{Introduction}\label{intro}

It is generally accepted that present-day spheroidal galaxies host
supermassive black holes (BHs) at their centres (\citealp{Magorrian98,
  Genzel99}). In addition, strong correlations are found between
BH masses and properties of their host galaxies
(\citealp{Ferrarese00, Gebhardt00, Tremaine02, Haering04, Graham07, 
  Gueltekin09, Burkert10, McConnell13}) as the bulge mass, the stellar
velocity dispersion, the Sersic index or the number of globular
clusters. This if often interpreted as an evidence for a co-evolution
between the spheroidal component of host galaxies and their BHs (see
\citet{Kormendy13} for a recent review). However, these relations were
also shown to possibly have a ``non-causal'' origin and thus, may
be just a consequence of statistical merging processes
(\citealp{Pengrandom, Hirschmann10, Jahnke11}). In addition, some
recent observations of BH growth and star formation in individual
objects appear to contradict the picture of a simple one-to-one
co-evolution over time (e.g. \citealp{Mullaney12, Bongiorno12,
  Rosario12, Hickox14}). Thus, the details of a connected growth of
BHs and their host galaxies remain an unresolved puzzle.  

During their lifetime, BHs are assumed to undergo
several episodes of significant gas accretion, during which this
accretion powers luminous quasars or active galactic nuclei (AGN)
(\citealp{Salpeter64, Zeldovich64,   Lynden69}). By estimating the
total energy radiated by AGN over their whole lifetime, it can be
shown that nearly all the mass seen in dormant BHs today can
be accumulated during the periods of observed bright AGN activity
(\citealp{Soltan82}). This implies that there is not much room left
for ``dark'' or obscured accretion.

Recent progress in detecting AGN (particularly faint and obscured
ones) was achieved by analysing data from X-ray surveys in the
hard and soft X-ray range (XMM-Newton, Chandra, ROSAT, ASCA, eRosita
e.g. \citealp{Miyaji00, LaFranca02, Cowie03, Fiore03, Barger03,
  Ueda03, Hasinger05, Barger05, Sazonov04, Nandra05, Ebrero09, Aird10,
  Fiore12, Kolodzig13}). Interestingly these studies revealed a
possibly at first sight puzzling behaviour of AGN concerning the
evolution of the co-moving number density of AGN: the number density
of successively more luminous AGN peaks at higher redshifts than the
one of less luminous AGN, with the lowest luminosity AGN showing an
almost constant co-moving number density. Making the simplified (and
naive) assumption that AGN luminosity (i.e. BH accretion) is
proportional to BH mass (which we would expect if BHs
are accreting at the Eddington rate, $L \propto M_{\bullet}$) would
imply that very massive BHs seem to be already in place at
very early times, whereas less massive BHs seem to evolve
predominantly at lower redshifts. This behaviour is called
``downsizing'' or ``anti-hierarchical'' growth of BHs. The
downsizing trend is also seen in the optical (\citealp{Cristiani04,
  Croom04, Fan04, Hunt04, Richards06,  Wolf03}) and the near-infrared
(NIR) (e.g. \citealp{Matute06}). 

This observational result seems to be in conflict with the ``naive''
expectations arising from the currently favoured hierarchical
structure formation paradigm based on the Cold Dark Matter (CDM) model
(\citealp{Peebles65, White78,   Blumenthal85}). In this framework, low
mass halos are expected to form first and more massive halos to grow
later over time via subsequent merging and smooth accretion. However,
it is now well known that the evolutionary history of observable
galaxies also follows an anti-hierarchical or downsizing behaviour,
with several independent observational indicators suggesting that
more massive galaxies formed their stars and had their star formation
quenched earlier than low-mass galaxies, which continue forming
stars to the present day  (see \citet{Fontanot09} for an overview).

To theoretically explore the formation and evolution of BHs and AGN
and its (possible) connection to the evolution of their host galaxies,
a large amount of studies were published based on semi-analytic models
of  galaxy formation within which mechanisms for BH formation and
evolution were included (e.g. \citealp{Kauffmann00, Volonteri03,
  Granato04, Menci04, Bromley04, Monaco05, Croton06, Bower06,
  Marulli08, Somerville08, Bonoli09, Fanidakis12, Hirschmann12,
  Neistein13, Menci13}). It was shown by some of the above studies
that an anti-hierarchical behaviour can be explained within a
hierarchical structure formation scenario (\citealp{Marulli08,
  Bonoli09, Fanidakis12, Hirschmann12, Menci13}), when considering
different models for BH accretion and AGN triggering and/or accounting
for dust obscuration. Unfortunately, these different models have not
been able to draw a completely uniform picture for the origin of the
downsizing behaviour as they differ in many details.         

In addition to these semi-analytic models, also a large number of
other studies investigated the predictions of the cosmological
$\Lambda$CDM model for the formation and evolution of super-massive
BHs (SMBH) and AGN. Some works made predictions based on 
nearly purely analytic models (\citealp{Efstathiou88, Haehnelt93,
  Haiman98}) and on semi-empirical models  (\citealp{Shankar09,
  Shankar10a, Shankar12}). 

Several years ago, numerical hydrodynamic simulations of isolated
galaxy mergers started to include prescriptions for BH growth and AGN
feedback using ``sub-grid'' recipes, where BH accretion according to
the Bondi-Hoyle-Littleton formula was assumed (\citealp{Springel05b,
  Hopkins06, DiMatteo05, Robertson06c, Li07, Sijacki07, Johansson09_1,
  Fabjan10, Barai13, Choi13} and references therein). Feedback from AGN, 
however was, implemented in different ways, ranging
from a purely thermal energy injection (e.g. \citealp{Springel05b}) and 
a further distinction between radio- and quasar mode 
(by injecting bubbles, e.g. \citealp{Sijacki07}, or just upscaling
of the feedback efficiencies, e.g. \citealp{Fabjan10}) to mechanical
momentum-driven winds (e.g. \citealp{Choi13}).

To assess \textit{statistical} properties of BHs and AGN in a full
cosmological context and in a more self-consistent way than it can be
achieved by e.g. semi-analytical or semi-empirical models,
\textit{large cosmological}, hydrodynamic simulations were performed
by several recent studies (\citealp{DiMatteo08, McCarthy10, Degraf10,
  McCarthy11, Degraf11, Booth11, DiMatteo12, Degraf12, Haas13,
  Rosas-Guevara13, Khandai14}) including BH growth and feedback from
AGN. Compared to semi-analytical models, they have the advantage that
the dynamics of the baryonic component (gas and stars) and the
interaction between baryonic matter can be followed directly in a
cosmological context, even if the spatial and mass resolution, at
present, is not high enough to accurately simulate small scale
physical processes like e.g. the formation of stars and BHs with the
associated feedback. Instead, they have to be computed in a simplified
manner with sub-grid/sub-resolution models. Nevertheless, they provide
a powerful tool to assess the cosmic evolution of statistical AGN and
BH properties and thus, go clearly beyond the hydrodynamic isolated
merger simulations and/or semi-analytic models.   

\citet{DiMatteo08} and \citet{Booth11}, for example, investigated the
global history of black hole mass assembly and the evolution of the
BH-bulge mass relations in cosmological simulations with a maximum box
size of $\sim (50\ \mathrm{Mpc})^3$. Such simulations have also been
successful in reproducing the low-luminosity end of the  AGN
luminosity function (\citealp{Degraf10}) or BH clustering properties
(\citealp{Degraf11}) but have been limited by a box size being not
large enough to follow the evolution of luminous AGN with
$L_{\mathrm{bol}} \geq 10^{45}$~erg/s (i.e. AGN luminosities above the
exponential cut-off of the luminosity function). In a recent
  study by \citet{Khandai14} they have used a larger cosmological
  volume of $(100\ \mathrm{Mpc/h})^3$ (where the luminous end is also
  accessible), but they typically underestimated luminous AGN
  at $z>0.5$ and overestimated them at $z<0.5$. Focusing 
only on high redshifts, \citet{DiMatteo12} and \citet{Degraf12}
investigated the growth of the first quasars appearing in the Universe
by performing cosmological simulations with a large box size of $(500\
\mathrm{Mpc/h})^3$, however, only run down to redshift $z=5$ finding a
rapid, early growth of BHs driven by cold, infalling gas flows.   

In this work, we extend their studies of the cosmic evolution of
statistical properties of BH growth and AGN evolution at
\textit{lower} redshifts, from $z=5$ down to $z=0$. We analyse a
subset of cosmological, hydrodynamical simulations from the
\textit{Magneticum Pathfinder} simulation set (Dolag et al., in prep),
which includes {\textit{a very large cosmological box providing a
    volume of ($500$~Mpc)$^3$}}. The simulations include a
self-consistent model for BH growth using the Gadget3 code containing
several modifications for modelling the growth of BHs with respect to
previous studies (e.g. \citealp{DiMatteo08, DiMatteo12, Khandai14}). 

In the course of this study, we will particularly focus on the
evolution of the AGN luminosity function and the connected
\textit{anti-hierarchical/downsizing} trend in AGN number density
evolution with the aim to understand its origin within a
\textit{hierarchical} structure formation scenario. In this context,
we test the effect of empirically motivated dust obscuration models
and discuss implications of the downsizing trend on the interplay
between AGN luminosities, BH mass and Eddington-ratios. {Overall,
  in this study, we go beyond previous papers of
  e.g. \citet{DiMatteo08}, \citet{Degraf10} or \citet{Khandai14} as we
  examine the evolution of the AGN luminosity function in a by a
  factor of $\sim 45$ larger cosmological volume  with comparably good
  (even if slightly lower than in \citealp{Khandai14}) resolution
  allowing us to probe particularly 
  luminous AGN with $L_{\mathrm{bol}} \geq 10^{45}$~erg/s. In
  addition, the simulation code is improved in various details, 
  regarding the SPH implementation used as well as the richness
  of physical processes included, including details of the handling
  of the BH sink particles. Here the most noticeable difference is
  reflected in our ability to follow BHs in galaxies which are inside
  of galaxy clusters.} 

This study is organised as follows. Section \ref{simulation} gives an
overview of the simulation code and the corresponding model for black
hole growth and AGN feedback adopted in our simulations. In Section
\ref{obs} we present some basic properties of present-day and
high-redshift BHs and galaxies, which are compared to
observational data. {We study the evolution of the AGN luminosity
function in Section \ref{AGNlum_lowz} at ($z=0-5$)}, considering
bolometric, soft and hard X-ray and (for $z=0$ only) also radio
luminosities, and compare our results to observational data. In
Section \ref{Anti-hier}, we explicitly show the evolution of the AGN
number density and discuss the anti-hierarchical trend in BH
growth, its consequences on the connection between BH masses
and AGN luminosities and its numerical origin in our simulations. In
the end, in Section \ref{downdis}, we summarise and discuss our main
results.

\section{The simulations}\label{simulation} 

In this paper, we analyse a subset of cosmological boxes from the 
Magneticum Pathfinder simulation set (Dolag et al., in prep.).
{From this set, we selected a large size cosmological simulation
  with a co-moving box size of $(500 \mathrm{Mpc})^3$,  simulated
  with an initial particle number of $2 \times 1,564^3$ (500Mpc/hr,
  which is our large ``high-resolution'' run throughout this study)
and one cosmological simulation with a  smaller co-moving box size of
$(68 \mathrm{Mpc})^3$ (only run down to $z=1$), with a particle
number of $2 \times 576^3$ and thus, an increased (``ultra-high'')
resolution (68Mpc/uhr).} Table \ref{tab_simruns} gives an overview of
the two simulation runs. We have adopted a $\Lambda$CDM model with
parameters chosen according to the seven-year  \textit{Wilkinson
  Microwave Anisotropy Probe} WMAP7 data (\citealp{Komatsu11}) with 
$\Omega_{\mathrm{m}}= 0.272$, $\Omega_{\mathrm{b}} = 0.0456$
$\Omega_\Lambda= 0.728$ and $h = 0.704$. The initial power spectrum
follows $n = 0.963$ and is normalised to $\sigma_8 = 0.809$.

\begin{table*}
\centering
\caption{Overview of the two simulation runs which are analysed in
  this study.}
\begin{tabular}{p{1.5cm} p{1.5cm} p{2.cm} p{2.5cm} p{1.5cm} p{1.5cm}
    p{1.5cm} p{2.5cm}}\hline\\ 
  \bf{Name} &
  \bf{Box size} & \bf{Resolution level} & \bf{Initial particle number} &
  \bf{m(dm)} & \bf{m(gas)} & \bf{m(stars)} & \bf{Softening length (dm,gas,stars)} \\
 & [Mpc/h] &  &  & [$M_\odot/h$] & [$M_\odot/h$] & [$M_\odot/h$] & [kpc/h]
 \\ \hline \hline 
500Mpc/hr & 352 & hr &$ 2 \times 1,564^3$ & $6.9 \times 10^8
$  & $1.4  \times 10^8 $ & $3.5  \times 10^7 $ & 3.75, 3.75, 2.0\\
68Mpc/uhr & 48 & uhr &$ 2 \times 576^3$ & $3.6 \times 10^7
$ & $7.3 \times 10^6 $ & $1.8 \times 10^6 $ & 1.4, 1.4, 0.7 \\ 
\hline
\end{tabular}
\label{tab_simruns}
\end{table*}

\subsection{The numerical code}\label{simcode}

{Our simulations are based on the parallel cosmological TreePM-SPH 
code {\small P-GADGET3} (\citealp{Springel05gad}). The code uses an
entropy-conserving formulation of SPH \citep{2002MNRAS.333..649S}. 
We are using a higher order kernel based on the bias-corrected, 
sixth-order Wendland kernel \citep{2012MNRAS.425.1068D} with 295 
neighbours which together with the usage of a low-viscosity SPH scheme
allows us to properly track turbulence within galaxy clusters 
\citep{2005MNRAS.364..753D,2013MNRAS.429.3564D}. We also allow for
isotropic thermal conduction with $1/20$ of the classical Spitzer value
\citep{2004ApJ...606L..97D}. The simulation code includes a treatment
of radiative cooling, heating from a uniform time-dependent ultraviolet
background and star formation with the associated feedback
processes. The latter is based on a sub-resolution model for the
multiphase structure of the interstellar medium \citep{Springel03}.}

Radiative cooling rates are computed by following the same procedure 
presented by \citet{Wiersma09}. We account for the presence of the
cosmic microwave background (CMB) and of ultraviolet (UV)/X-ray
background radiation from quasars and galaxies, as computed by
\citet{Haardt01}. The contributions to cooling from each one of 11
elements (H, He, C, N, O, Ne, Mg, Si, S, Ca, Fe) have been
pre-computed using the publicly available CLOUDY photo-ionisation code   
(\citealp{Ferland98}) for an optically thin gas in (photo-)ionisation
equilibrium.  

\begin{figure}
\begin{center}
  \epsfig{file=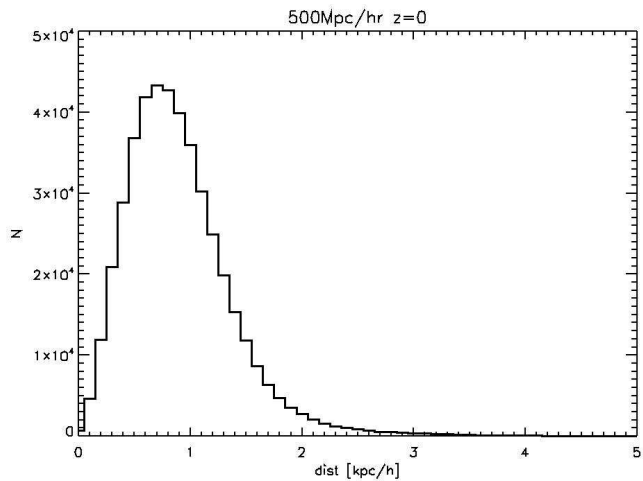,
    width=0.5\textwidth} 
   \caption{{Histogram of the distances of the black holes to their
       halo potential minimum in the 500Mpc/hr simulation. All BHs are
     maximal 2~kpc away from the potential minium which is reasonable
     given a softening length of 5.2~kpc in this run.}}  {\label{dist}} 
\end{center}
\end{figure}

In the multiphase model for star-formation \citep{Springel03}, the ISM
is treated as a two-phase medium where clouds of cold gas form from
cooling of hot gas and are embedded in the hot gas phase assuming
pressure equilibrium whenever gas particles are above a given
threshold density. The hot gas within the multiphase model is heated
by supernovae and can evaporate the cold clouds. A certain fraction of
massive stars (10 per cent) is assumed to explode as supernovae type
II (SNII). The released energy by SNII ($10^{51}$~erg) is modelled to
trigger galactic winds with a mass loading rate being proportional to
the star formation rate (SFR) to obtain a resulting wind velocity of
$v_{\mathrm{wind}} = 350$~km/s. 

Our simulations also include a detailed model of chemical evolution
according to \citet{Tornatore07}. Metals are produced by SNII, by
supernovae type Ia (SNIa) and by intermediate and low-mass stars in
the asymptotic giant branch (AGB). Metals and energy are released by
stars of different mass to properly account for mass-dependent
life-times (with a lifetime function according to
\citealp{Padovani93}), the metallicity-dependent stellar yields by
\citet{Woosley95} for SNII, the yields by \citet{vandenHoek97} for AGB
stars and the yields by \citet{Thielemann03} for SNIa. Stars of
different mass are initially distributed according to a Chabrier
initial mass function  (IMF; \citealp{Chabrier03}).

\begin{figure*}
\begin{center}
  \epsfig{file=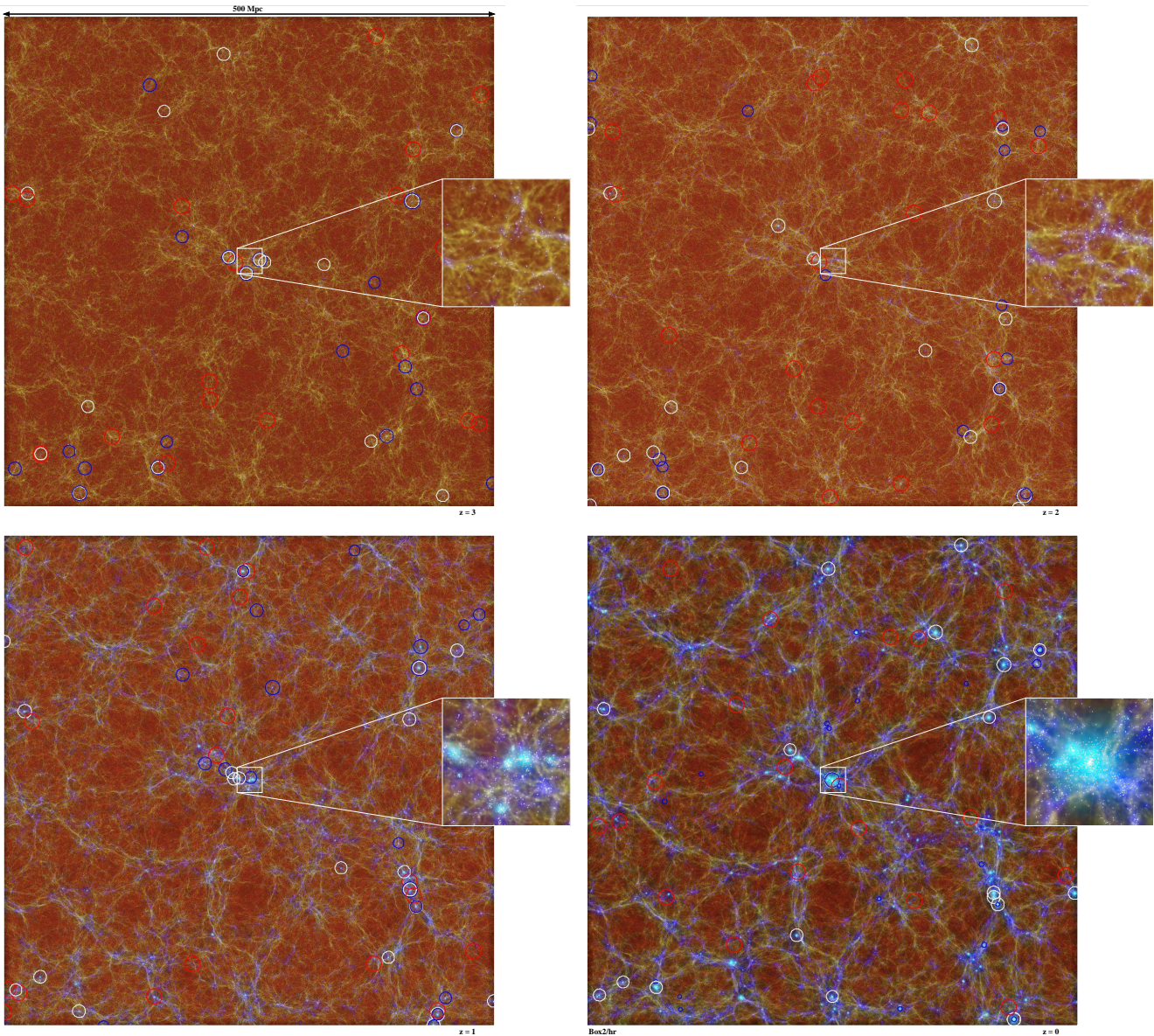,
    width=1.0\textwidth} 
   \caption{Shown is a 500 Mpc wide, 70 Mpc thick slice through the
       cosmological baryonic mass distribution (stellar and gaseous
       density) of the Box2/hr simulation at different redshift steps
       ($z=3,2,1$ and 0). This is the result of a ray tracing
       visualisation using {\em SPLOTCH}
       \citep{2008NJPh...10l5006D}. White, blue and red circles
       indicate the 20 BHs within this slice which have the highest
       masses, the highest  Eddington-ratios and the highest accretion
       rates, i.e. AGN luminosities, respectively. The sizes of the
       circles are scaled logarithmically with the different values,
       normalised to the maximum value of each quantity. A zoom onto a
       region where the most massive cluster forms is shown, together
       with all galaxies having a stellar mass larger than
       $10^{10}M_\odot$ illustrated by white crosses, whereas the
       white diamonds show all BHs.}  {\label{Sim_vis}} 
\end{center}
\end{figure*}

\begin{figure*}
\begin{center}
  \epsfig{file=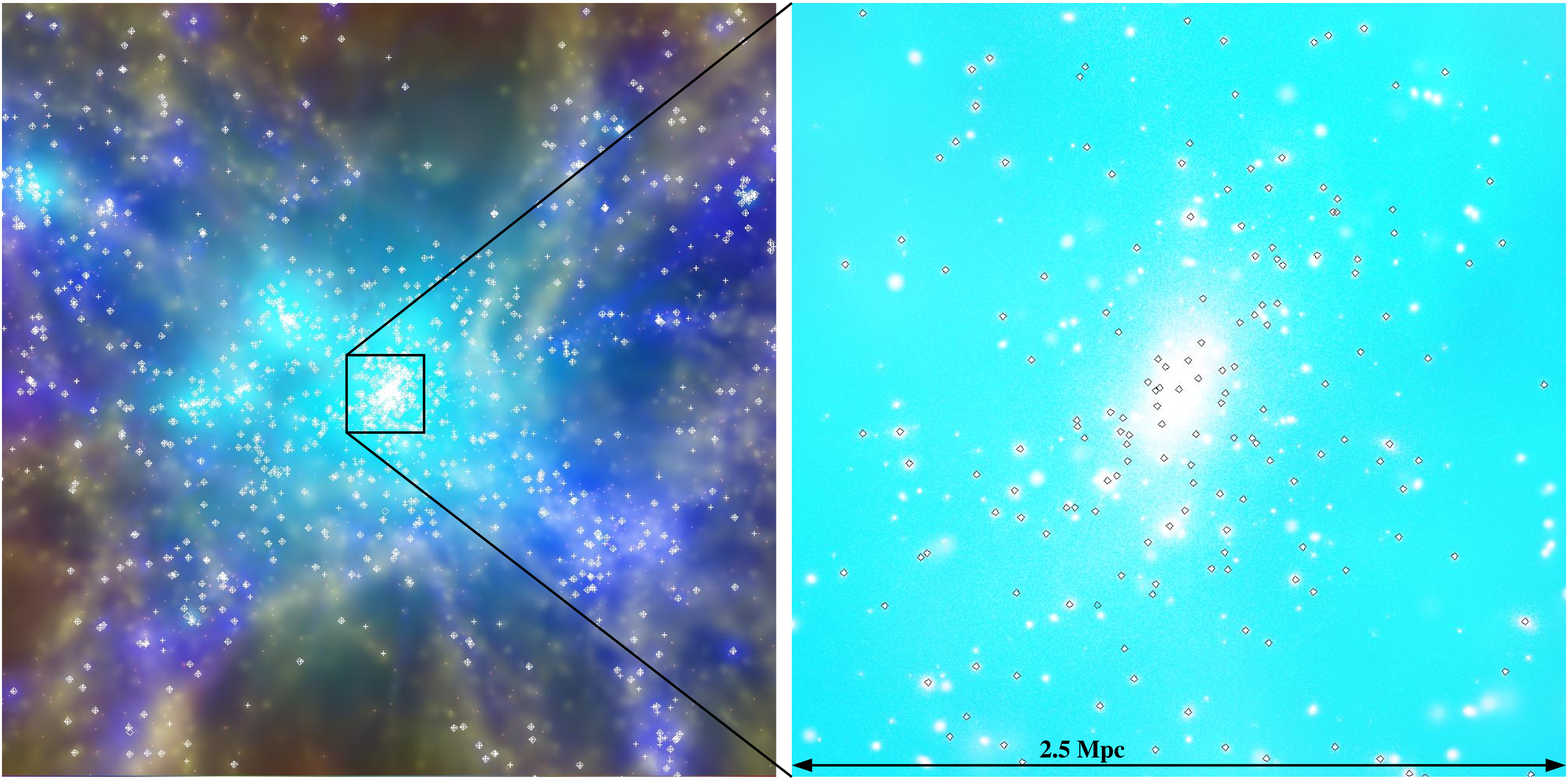,width=1.0\textwidth}
   \caption{{Shown is the 25 Mpc wide zoom onto the galaxy cluster cluster
            at z=0, where analogous to figure \ref{Sim_vis} all galaxies with
            stellar mass larger than $10^{10}M_\odot$ are shown as white crosses, 
            and all BHs are shown as white diamonds. The right panel visualises
            a further zoom into the cluster. The region shown is 2.5 Mpc wide
            and correspond to roughly one third of the virial size of the cluster.
            In the ray tracing visualisation, the white colours reflect the stellar
            component, while the light blue colours correspond to the hot phase of
            the ICM. Black diamonds mark all the BHs in the simulation.}
}  {\label{Sim_vis2}}
\end{center}
\end{figure*}

\subsection{The BH growth model}\label{BHgrowth}

Most importantly, our simulations also include a prescription for
BH growth and for a feedback from active galactic nuclei (AGN)  
based on the model presented in \citet{Springel05a} and
\citet{DiMatteo05} including the same modifications as in the study of  
\citet{Fabjan10} and {some new, minor changes for BH seeding and
  BH ``pinning'' which are explained in later in this section. }

As for star formation, the accretion onto BHs and the
associated feedback adopts a sub-resolution model. BHs are
represented by  collision-less ``sink particles'' that can grow in
mass by accreting gas from their environments, or by merging with
other BHs.  

The gas accretion rate $\dot{M}_\bullet$ is estimated by using the
Bondi-Hoyle-Lyttleton approximation (\citealp{Hoyle39, Bondi44, 
  Bondi52}): 
\begin{equation}\label{Bondi}
\dot{M}_\bullet = \frac{4 \pi G^2 M_\bullet^2 \alpha
  \rho}{(c_s^2 + v^2)^{3/2}}, 
\end{equation}
where $\rho$ and $c_s$ are the density and the sound speed of the 
surrounding (ISM) gas, respectively, $v$ is the velocity of the black
hole relative to the surrounding gas and $\alpha$ is a boost factor
for the density and the sound speed which typically is set to $100$ as
in most related works (unless a more detailed description as
introduced in \citet{2009MNRAS.398...53B} is used) and accounts for
the fact that in cosmological simulations we can not resolve the
intra-cluster medium (ICM) properties within the vicinity of the
BH. The BH accretion is always limited to the Eddington rate (maximum
possible accretion for balance between inwards directed gravitational
force and outwards directed radiation pressure): $\dot{M}_\bullet =
\min(\dot{M}_\bullet, \dot{M}_{\mathrm{edd}})$. Note that the detailed
accretion flows onto the BHs are unresolved, we can only capture BH
growth due to the larger scale gas distribution, which is resolved. 

Once the accretion rate is computed for each black hole particle the
mass continuously grows. To model the loss of this accreted gas from
the gas particles, a stochastic criterion is used to select the
surrounding gas particles to be accreted. Unlike in
\citet{Springel05a}, in which a selected gas particle contributes to
accretion with \textit{all} its mass, we include the possibility for a gas 
particle to accrete \textit{only with a slice of its mass}, which
corresponds to 1/4 of its original mass. This way, each gas particle
can contribute with up to four generations of BH accretion events,
thus providing a more continuous description of the accretion process.  

The {total released energy $\dot{E}$ is related to the BH
accretion rate by}
\begin{equation}\label{Lrad}
\dot{E} = \epsilon_r \dot{M}_\bullet c^2,  
\end{equation} 
{where $\epsilon_{\mathrm{r}}$ is the radiative efficiency, for which
we adopt a fixed value of 0.2. Here we are using a slightly larger 
value than the one standardly assumed (=0.1 ) for a radiatively
efficient accretion disk onto a non-rapidly spinning BH according to
\citealp{Shakura73}, (see also  \citealp{Springel05gad,
  DiMatteo05}). This choice is motivated by observations of 
\citet{2011ApJ...728...98D} who find higher $\epsilon_r$ for higher BH 
masses and by our numerical
resolution. }

{We assume that a fraction $\epsilon_{\mathrm{f}}$ of this energy is 
thermally coupled to the surrounding gas so that 
\begin{equation}
\dot{E}_{\mathrm{f}} = \epsilon_{\mathrm{r}}\epsilon_{\mathrm{f}}
\dot{M}_\bullet c^2 
\end{equation} 
is the rate of the energy feedback. $\epsilon_{\mathrm{f}}$ is a free
parameter and typically set to $0.15$ (as usually done in simulations
which follow the metal depending cooling function, see for example
\citealp{Booth11}).} The energy is distributed kernel weighted to the 
surrounding gas particles in an SPH like manner. 

Additionally, we incorporated the feedback prescription according to
\citet{Fabjan10}: we account for a transition from a quasar- to a
radio-mode feedback (see also \citealp{Sijacki07}) whenever the
accretion rate falls below an Eddington-ratio of $f_{\mathrm{edd}} :=
\dot{M}_{\mathrm{r}}/ \dot{M}_{\mathrm{edd}} < 10^{-2}$. During this
radio-mode feedback we assume a 4 times larger feedback efficiency
than in the quasar mode. This way, we want to account for massive BHs,
which are  radiatively inefficient (having low accretion rates), but
which are efficient in heating the ICM by inflating hot bubbles in
correspondence of the termination of AGN jets. The total
  efficiency in the radio mode is very close to the value of 0.1
  ($=0.15 \times 0.2 \times4$). This is the canonical value, which
  \citet{Churazov05} estimated to be needed to balance cooling by AGN
  feedback. 

Note that we also, in contrast to \citet{Springel05a}, modify the mass
growth of the black hole by taking into account the feedback, e.g. $
\Delta M_\bullet = (1-\epsilon_{r})\dot{M}_\bullet \Delta
t$. Furthermore, we introduced some additional, technical
modifications of the original implementation which we will now
summarise:  

\noindent
{\bf (I)} One difference with respect to the original implementation by
\citet{Springel05a} concerns the seeding of BH particles. In the
implementation by \citet{Springel05a}, BH particles are seeded in a
halo whenever it first reaches a minimum (total) friends-of-friends
(FoF) halo mass, where the FoF is performed on the dark matter
particles only. In order to guarantee that BHs are seeded only in
halos {representing clearly resolved galaxies}, where sufficient
star formation took place, our implementation performs a FoF algorithm 
on star particles, grouping them with a linking length of 0.05
times the mean separation of the DM particles\footnote{Note that this
  linking length is thus much smaller than that, $0.15 - 0.20$,
  originally used, to identify virialised halos.}. 

In the ``hr'' simulation presented here, a total stellar mass of roughly
$10^{10} M_\odot/h$ is needed (corresponding to a couple of
  hundreds of star particles) for a halo to be seeded with a BH
particle (starting with a seed mass of $3.2\times10^5 M_\odot/h$).  
In the ``uhr'' simulation we are using slightly smaller values due to
the better underlying resolution (BH seed masses of $8\times10^4
M_\odot/h$ in galaxies with a minimum stellar mass of $2.5\times
10^{9} M_\odot/h$). While the BH then grows very fast until it reaches
the stellar-BH-mass relation, this recovers the BH feedback within the
galaxies which would have been present if resolution had allowed to
seed BHs earlier. This also avoids to imprint any stellar-BH-mass
relation from the beginning. Finally, we choose the seeded BHs at the
position of the star particle with the largest binding energy within
the FoF group, instead of at the dark matter particle with the maximum
density, as originally implemented.

\noindent
{\bf (II)} In the original implementation by \citet{Springel05a},
black holes are forced to remain within the host galaxy by pinning
them to the position of the particle found having the minimum value of
the potential among all the particles lying within the SPH smoothing
length computed at the BH position. Within a cosmological
context an aside effect of this criterion is that, due to the
relatively large values of SPH smoothing lengths, a BH can be
removed from the host galaxy whenever it becomes a satellite, and is
spuriously merged into the BH hosted by the central halo
galaxy. We have relaxed this criterion and do not apply any pinning of
the BH particles to the minimum potential within the smoothing
length. 

To avoid that the BH particles are wandering away from the centre of
galaxies by numerical effects, we take several measures, in addition
to the original implementation of the BH treatment: first, we enforce
a more strict momentum conservation within the implementation of gas
accretion by forcing momentum conservation for the smooth accretion of
the gas and then do not model any  momentum transfer when swallowing
gas\footnote{Note that otherwise one would statistically account for
  the momentum transfer of accreted gas twice.}. Additionally we
implemented the conservation of momentum and centre of mass when two
BH particles are {merging\footnote{Note that the original scheme
    the merged BH have had the position and velocity of the BH with
    the smaller particle ID.}}.  

\noindent
{\bf (III)} Moreover, in contrast to the original implementation, we
have explicitly included a dynamical friction force, {which is
  switched on unless the underlying simulation has a high enough
  resolution so that the cosmological simulations can numerically
  resolve dynamical friction reasonably well}. To estimate the typical
friction force induced onto a BH particle we are using the following
approximation of the Chandrasekhar formula
(\citealp{Chandrasekhar43}): 
\begin{equation}
   F_\mathrm{df} = -4 \pi \left(\frac{G
       M_\bullet}{v}\right)^2\rho\;\mathrm{ln}(\Lambda)
   \left(err(x)-\frac{2x}{\sqrt{\pi}}e^{-x^2}  
   \right)\frac{\vec{v}}{v}, 
\end{equation}
where $G$ is the gravitational constant and $M_\bullet$ is the mass of 
the BH. The local density $\rho$ in the vicinity of the black
hole as well as for the relative velocity $\vec{v}$ is calculated
using only the stellar and the dark matter components around the black
hole. The Coulomb logarithm is calculated as   
\begin{equation}
\mathrm{ln}(\Lambda) = \mathrm{ln}\left(\frac{R v}{G M_\bullet}\right)
\end{equation}
and $x=v\sqrt(2)/\sigma$, where we estimate $\sigma$ as one third of  
the maximum circular velocity of the hosting sub-halo and for $R$ (as 
typical size of the system) we used the half mass radius of the
sub-halo, hosting the BH. The parameters of the hosting
sub-halo for each BH particle are updated every time when
SubFind (\citealp{Dolag09}) is executed on-the-fly.   

\begin{figure*}
\begin{center}
  \epsfig{file=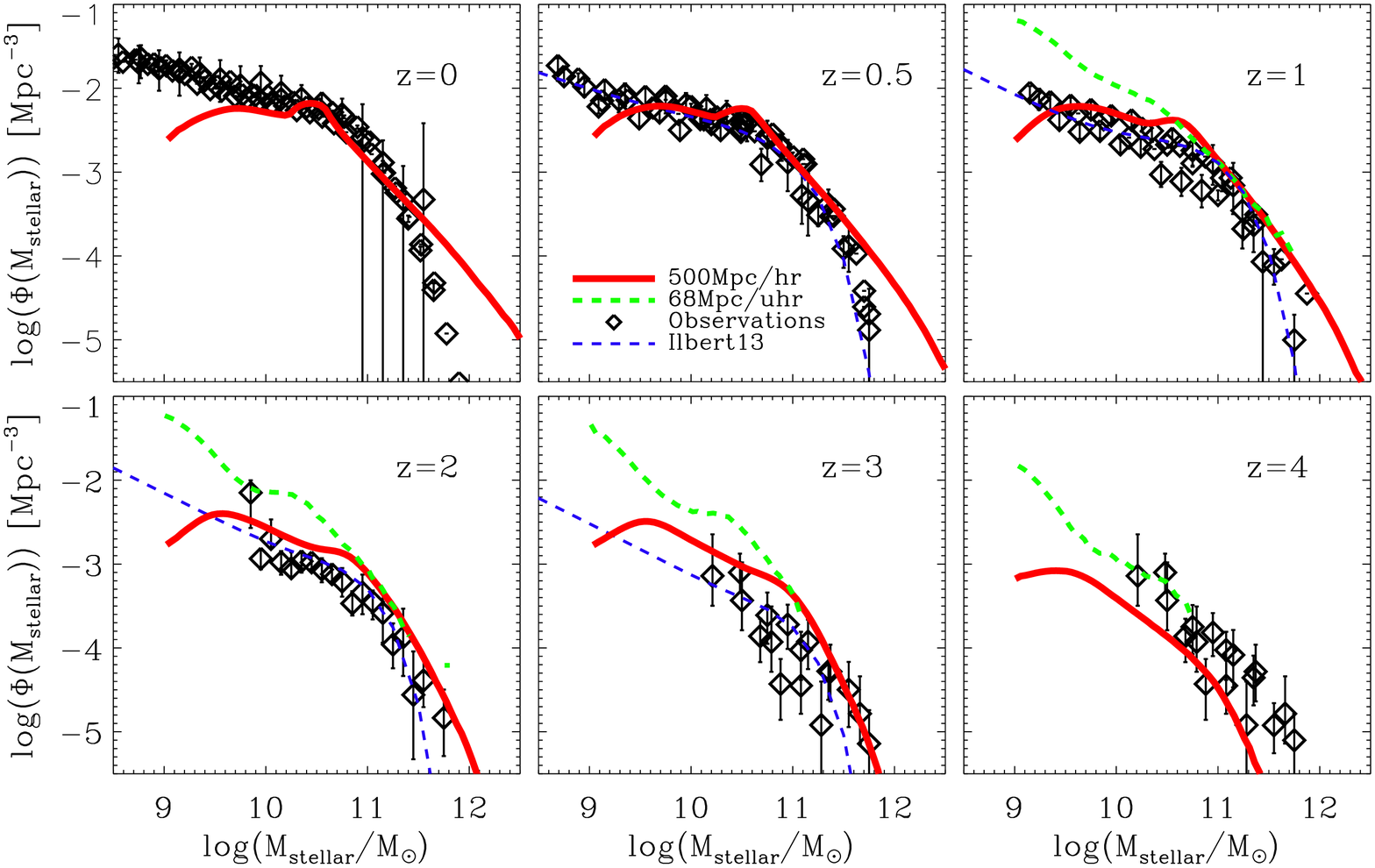,
    width=0.8\textwidth} 
  \caption{Evolution of the stellar mass function in the 500Mpc/hr 
    (red lines) and in the 68Mpc/uhr run (green dashed lines) compared
    to observational data (black symbols, \citealp{Perez08, Bundy05,
      Drory04, Fontana06,  Marchesini08, Ilbert10} and blue dashed
    lines, \citealp{Ilbert13}). At $z<1$ the massive end is
    over-estimated in the simulations due to a too inefficiently
    working radio-mode feedback.} 
 {\label{Evolstellar}}  
\end{center}
\end{figure*}

{This way a BH particle remains within the host galaxy, even
if it becomes a satellite of a larger halo and, compared to the
original scheme, we are able to track BHs also in satellite
galaxies in cluster environments. When the BHs are not placed
artificially on the minimum of the potential, of course, there is no
guarantee (due to numerical noise, 2 body scattering or when two
BHs are merging) that black hole particles are staying always exactly
at the local potential minimum. But due to the above handling of the
dynamical friction, with evolving time during the simulation, BHs sink
towards the minimum potential so that typical displacements from the
true potential minimum are smaller than the effective gravitational
softening. Therefore, they are orders of magnitude smaller than the
typical smoothing radius used for estimating the parameters in the
accretion model or used for distributing the feedback
energy. Fig. \ref{dist} shows explicitly a histogram for the distances
of BHs from the potential minimum in the 500Mpc/hr run. The
distribution is peaking at 0.7~kpc and the vast majority of BHs are
maximal 2~kpc away from the potential minimum which is reasonable for
the used smoothing length of 5.2~kpc.}  

Fig. \ref{Sim_vis} shows a visualisation (of the gas and stellar
mass density) of the 500Mpc/hr run at different redshift steps
($z=3-0$, different panels) focusing on a thin (70 Mpc thick)
slice. We have additionally indicated the 20 BHs with highest masses
(white circles), the highest Eddington-ratios ($f_{\mathrm{edd}} =
L/L_{\mathrm{edd}}$, blue circles) and the highest accretion rates
(and thus, luminosities, red circles). The sizes of the circles are
scaled logarithmically with the different values and normalised to the
maximum value of each quantity.

This visualises that AGN luminosity does not directly trace the mass
of a BH as AGN when selected by their BH mass (white) seem  
to be a better tracer of the underlying matter distribution, whereas
when selected by the Eddington-ratio (blue) or by their luminosity
(red) they are more located in less dense environments, what already
indicates the presence of a ``downsizing'' trend in their evolution.

{In particular, one can see that at $z=3$ many most massive BHs
  also have the highest accretion rates as the accretion
  rate/luminosity is at that time still largely related with BH mass
  (even if with some scatter, as discussed later in Fig. \ref{Lum_bh}).  
  With evolving time, this is not the case anymore and massive BHs
  accrete at a very broad range of Eddington-ratios so that at those
  later times, massive black holes hardly coincide with high
  Eddington-ratios and accretion rates.  In addition, at low
  redshifts, there are several objects with high Eddington-ratios
  located close to voids, these are low mass BHs (not necessarily
  satellites), which are growing by smooth gas accretion (as they do
  not have experienced much feedback).}  

{The inlay shows a zoom onto the region where the most massive
cluster forms today. Over-plotted here are all galaxies with stellar
masses above $10^{10}M_\odot$ (white crosses) and all BHs within this
slice of the  simulation (white diamonds). Fig. \ref{Sim_vis2} shows
a zoom onto the redshift zero slice (left panel) an a zoom onto the
central, 2.5~Mpc wide region of the cluster. Here, the stellar
populations of the galaxies (white) within our ray tracing
visualisation can be nicely seen contrasted to the light blue
ICM. Note that not all galaxies contain a BH as the initial seeding is
also related to the on-going SF in a galaxy. The position of all BHs
within this simulation are marked as black diamonds and nicely reflect
the ability of our implementation to keep the BHs at the centre of the
satellite galaxies (see also Fig. \ref{dist}).}

\section{Fundamental galaxy and BH properties and their
  evolution}\label{obs}  

In this section, we will discuss some fundamental properties of the 
simulated galaxies and BHs in the 500Mpc/hr run, as the stellar mass
function, the BH mass function and the BH-stellar mass relation in the
present-day Universe and at higher redshifts up to $z=4$. {To show
  the effect of resolution we additionally present the results for the
68Mpc/uhr run down to $z=1$.}

\subsection{The stellar mass function}\label{SMF}

Fig. \ref{Evolstellar} shows the stellar mass function at different
redshift steps as indicated in the legend. Simulation results (red solid
lines for the 500Mpc/hr run and green dashed lines for the 68Mpc/uhr run) are
compared with observational data from different studies (black
symbols: \citealp{Perez08, Bundy05, Drory04, Fontana06, Marchesini08,
  Ilbert10}; blue dashed lines: \citealp{Ilbert13}). {At $z=4$,
  the amount of galaxies in the 500Mpc/hr run is slightly
  under-estimated, which is a resolution effect: for 
the 68Mpc/uhr simulation the low-mass end of the stellar 
mass function at these high redshifts is consistent with the
observational data. Instead, down to $z=1$ the 500Mpc/hr simulation
results provide a reasonably good match with the observational data,
while the low-mass end in the 68Mpc/uhr run is over-estimated by up to 
1~dex.} 

{The over-estimation of low-mass galaxies is a well-known problem
  and most likely a consequence of our adopted model for stellar winds
  assuming a constant wind velocity for the ejected gas
  (e.g. \citealp{Oppenheimer06, Dave11}). It was repeatedly shown in
  literature that energy- or momentum-driven wind models can
  significantly reduce the baryon conversion efficiencies
  (e.g. \citealp{Hirschmann13} and references therein) and thus, also
  the low-mass end of the stellar mass function resulting in an improved
  match with the observational data (e.g. \citealp{Dave13,
    Puchwein13}). In these studies, energy-driven winds, for example,
  are shown to reduce the low mass end of the stellar mass function by
  at least 1~dex compared to a constant wind model which can account
  for the discrepancy between our 68Mpc/uhr simulation predictions and
  the observational constraints. In addition, models including ``early''
  stellar feedback (\citealp{Stinson13, Kannan14}) or radiation pressure
  (\citealp{Hopkins13}) seem to be particularly efficient in delaying
  star formation in low mass halos towards later times (i.e. in
  breaking the hierarchical formation of galaxies) and thus,
  predicting low baryon conversion efficiencies in these halos down to
  $z=0$ -- consistent with observational constraints.}  

Turning towards lower redshifts ($z<1$), the massive end of the
stellar mass function ($\log(M_{\mathrm{stellar}}/M_\odot) > 11$) is
significantly over-estimated in the 500Mpc/hr run, at $z=0$ by more 
than one order of magnitude for $10^{12} M_\odot$-mass
galaxies. {This is most likely a consequence of the thermal energy
  injection scheme in the ``radio-mode'' adopted in our model (see
  also \citealp{Puchwein13} who also over-estimate the massive end of
  the stellar mass function). Here we may speculate that a
  mechanical-momentum input from an AGN coupling to the ambient gas
  via a bipolar wind would be more efficient in limiting the infall
  and accretion onto the central BH and also star formation and thus,
  could help making elliptical galaxies "red and dead" by suppressing
  late star formation (\citealp{Choi14}). Such mechanisms are for
  example investigated  by \citet{Choi12, Debuhr12, Barai13, Dubois12}
  employing simulations of isolated galaxies or galaxy mergers.}

\begin{figure}
\begin{center}
  \epsfig{file=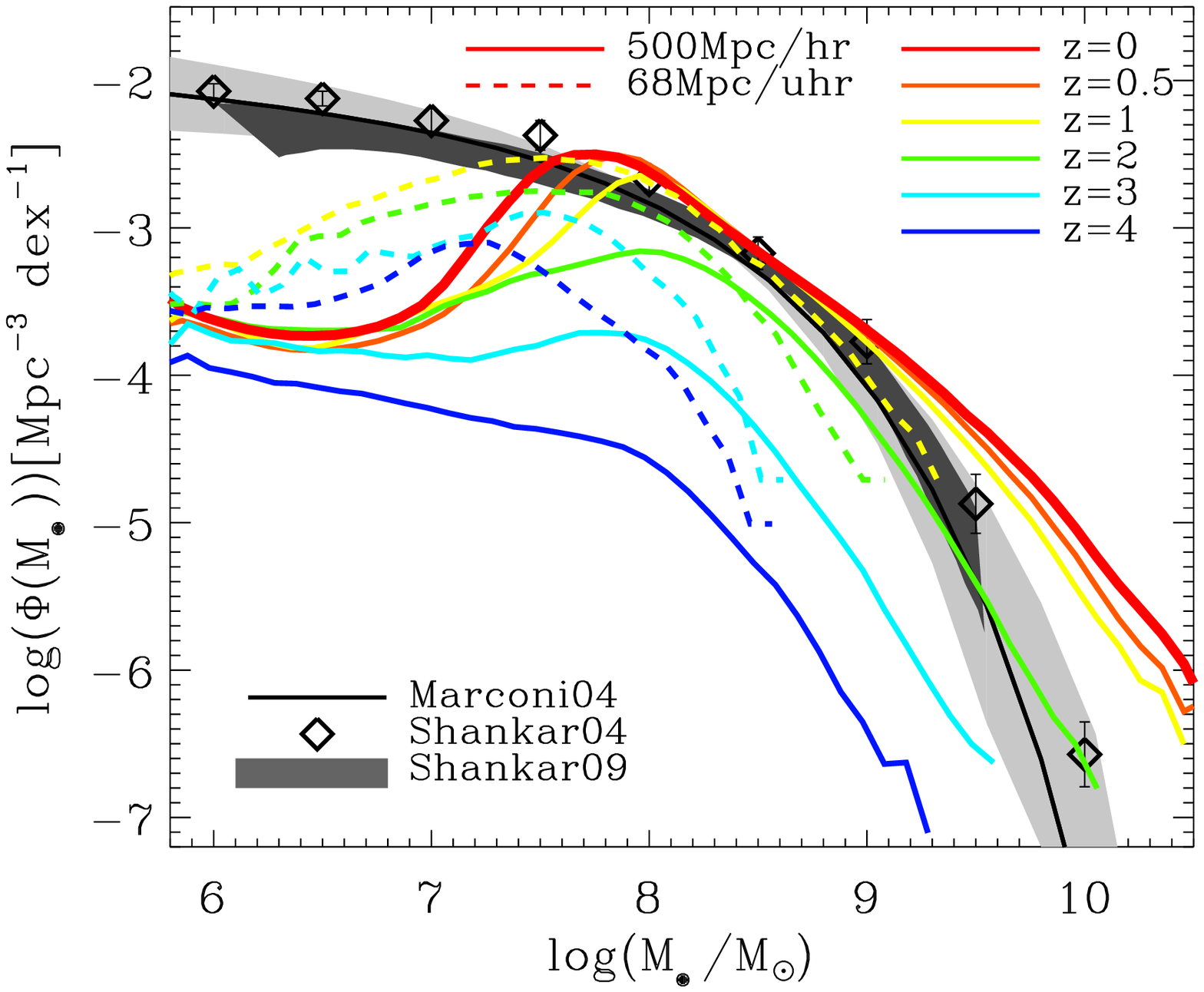,
    width=0.5\textwidth} 
  \caption{Evolution of the BH mass function in the 500Mpc/hr
    simulation (solid coloured lines) and in the 68Mpc/uhr simulation
    (dashed coloured lines). The present-day BH mass function (red
    line) is compared to observations from \citet{Marconi04,
      Shankar04} and \citet{Shankar09} (black lines and symbols with
    the grey shaded areas). We find a reasonable  agreement between
    observations and simulations for BH masses between $5\times
    10^7 M_\odot < M_\bullet < 3\times 10^9 M_\odot$, while above, the
    amount of massive BHs ($>3\times 10^9 M_\odot$) is
    over-estimated by  up to 1~dex (as a consequence of too
    inefficient radio-mode feedback). }  {\label{Evolbh}}   
\end{center}
\end{figure}

\begin{figure}
\begin{center}
  \epsfig{file=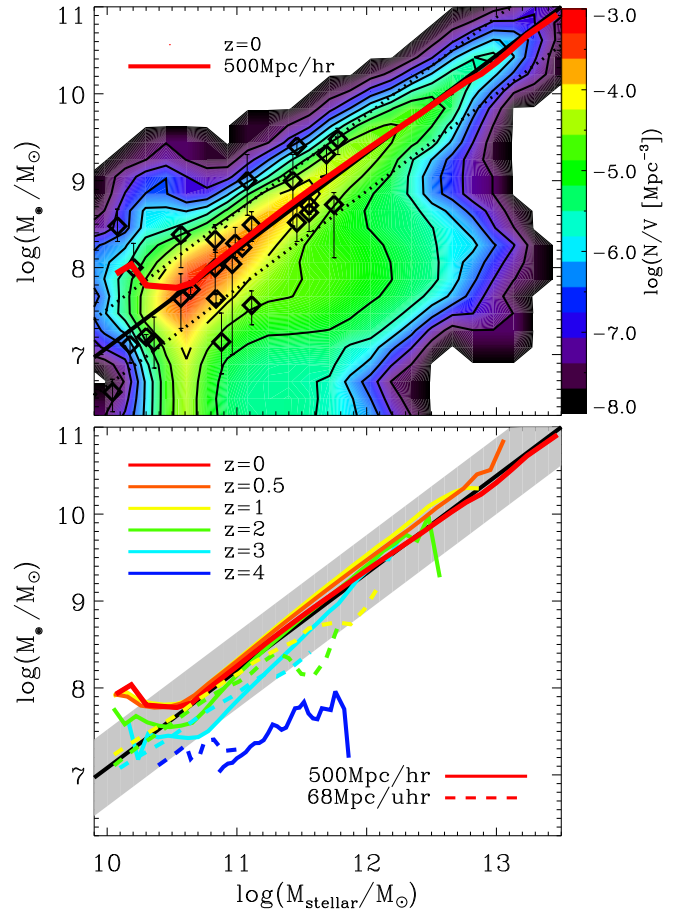,
    width=0.5\textwidth} 
  \caption{\textit{Upper panel}: present-day relation between black
    hole and stellar mass. Colour-coded contours show the simulated
    number density, the red line illustrates the mean for the
    500Mpc/hr run. The simulated relation perfectly matches the
    observational data (symbols illustrate their data set of
    \citet{Haering04}, the black solid line shows the fit to the data
    and the black dashed lines the corresponding 1-$\sigma$
    scatter). \textit{Bottom panel}: Evolution of the black
    hole-stellar mass relation in the 500Mpc/hr (solid lines) and the
    68Mpc/uhr run (dashed lines). For comparison, the observed
    present-day relation is also illustrated as in the upper panel. In
    the simulations, the BH-stellar mass relation is in place at
    $z=3$ and hardly evolves afterwards.} 
 {\label{BHbulgerel}}
\end{center}
\end{figure}

\begin{figure}
\begin{center}
  \epsfig{file=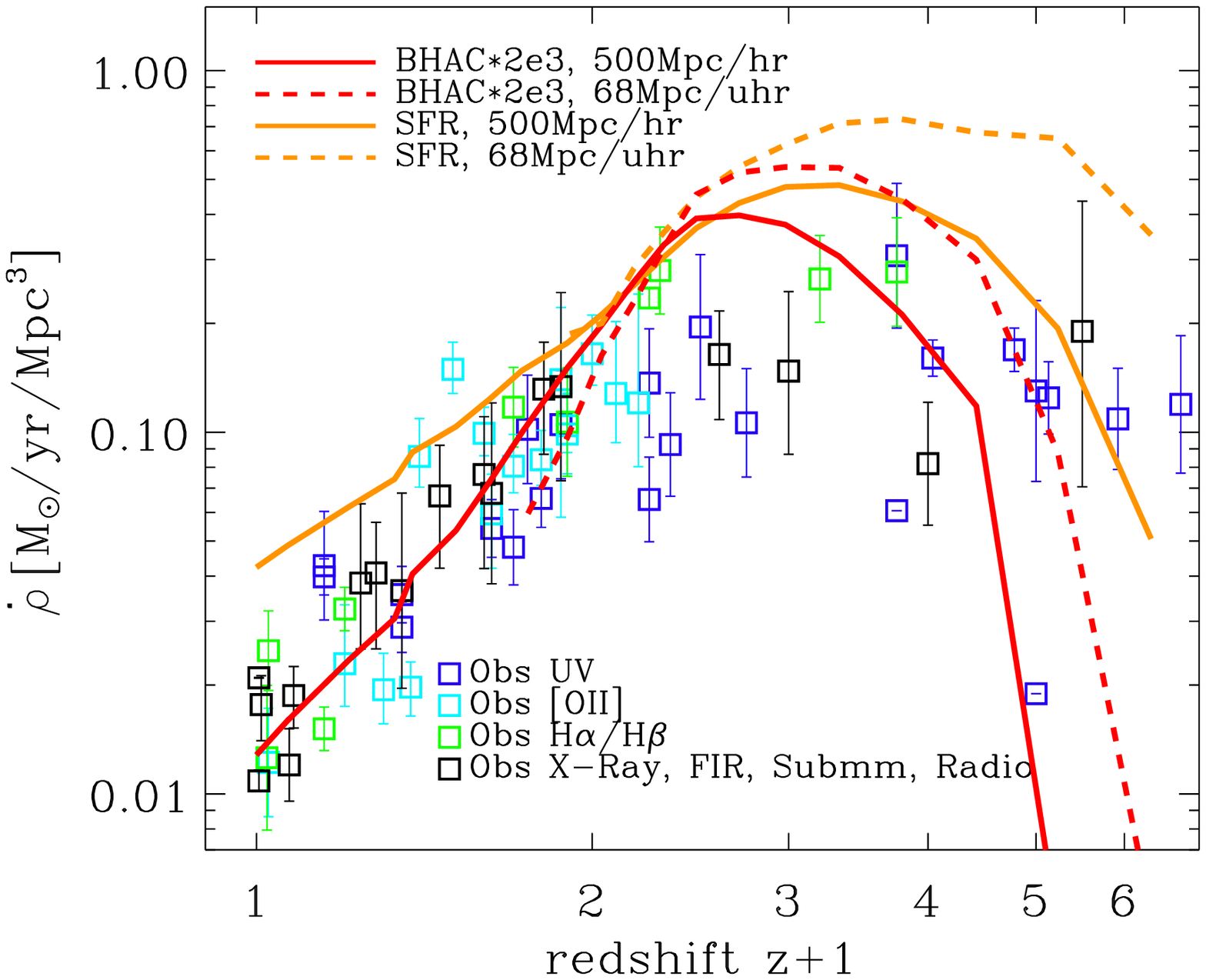, width=0.5\textwidth}
  \caption{Cosmic evolution of mean star formation (orange lines) and 
    BH accretion rate densities multiplied by a factor of
    $2\times 10^3$ (red lines) in the 500Mpc/hr (solid lines) and the
    68Mpc/uhr run (dashed lines). Both, star formation and BH
    accretion rate densities peak between $z \approx 1-2$ followed by
    a decline towards lower redshifts. Compared to observed cosmic
    star formation rate densities derived from  different wavebands
    (coloured symbols, \citealp{HopkinsA06}), simulations predict too
    high star formation rates at $z<1$ resulting in too massive
    galaxies (see Fig. \ref{Evolstellar}).}   
          {\label{madauplot}}
\end{center}
\end{figure}

\subsection{The BH mass function}\label{BHMF}

In Fig. \ref{Evolbh}, the present-day BH mass function of the
500Mpc/hr run (red line) is compared to observations from
\citet{Marconi04, Shankar04} and \citet{Shankar09} (black lines and
symbols and grey shaded areas) and is found to be in reasonably good
agreement for a BH mass range of $7.5 < \log(M_\bullet/M_\odot) <
9.5$. Below, the simulation is under-predicting the amount of low
massive BHs by almost one order of magnitude. {This is mainly
  related to a combination of too low resolution (main effect, see
  dashed coloured lines indicating the 68Mpc/uhr predictions) and BH
  seed masses.} 

{At the high mass end, the amount of massive BHs
$\log(M_\bullet/M_\odot) > 9.5$ is significantly over-estimated by up
to 2~dex. The high mass end is mostly influenced by the choice for 
the parameter regulating the strength for AGN feedback, but hardly
dependent on the efficiency parameter for regulating the radio-mode
feedback. Although the latter is supposed to regulate the late time
star formation and BH growth, in the implementation of the thermal
injection scheme in the ``radio-mode'' adopted in our model, is
still not efficient enough to lower the high mass end of the BH
mass function (as discussed for the stellar mass function). Therefore,
AGN feedback at high redshift is much more efficient in suppressing
star formation also at late times by pre-heating the
environment. Instead of a purely thermal energy injection, we
speculate that a mechanical-momentum input from an AGN coupling to the 
ambient gas via a bipolar wind would be more efficient in limiting the
infall and accretion onto the central BH (e.g. \citealp{Choi12,
  Barai13}) and thus, reducing the massive end of the BH mass
function.}

Turning towards higher redshifts (see coloured solid lines in
Fig. \ref{Evolbh} indicating the 500Mpc/hr run), the number density at
a given BH mass shows a very rapid evolution until $z=1$. At that
redshift, the high mass end is mostly in place and only the amount of
BHs below $\log(M_\bullet/M_\odot) < 8$ further increases until
$z=0$. This implies that BHs grow most strongly until $z=1$ which is
also in agreement when considering the cosmic evolution 
of the total BH accretion rates, which peak around $z=2$  and
fall below $0.1\ M_\odot\ \mathrm{yr}^{-1}\ \mathrm{Mpc}^{-3}$ after
$z=1$ (see also Fig. \ref{madauplot}).     

{To demonstrate the effect of resolution, we also show the BH mass
  functions at $z=1-4$ for the 68Mpc/uhr run (dashed coloured
  lines). The low mass end ($\log(M_\bullet/M_\odot)<8$) is significantly
  increased by up to 1~dex compared to the 500Mpc/hr run as a
  consequence of a faster mass growth of low mass BHs in the higher
  resolution run. This is in line with the increased low-mass end of the
  stellar mass function in the 68Mpc/uhr run due to inefficient
  stellar feedback.} 
{Above $\log(M_\bullet/M_\odot)>8$, however, the 68Mpc/uhr
  predictions roughly {converge against} the lower resolution results of
  the 500Mpc/hr run.} 

{Compared to observational constraints for the evolution of the BH
mass function (derived from integrating the continuity equation,
e.g. \citealp{Merloni08}), our predictions are broadly consistent: also in the
observations the massive end of the BH mass function is hardly
evolving after $z=1$, while the low-mass end is still increasing. This
trend already reflects the anti-hierarchical behaviour we will discuss
in the course of this study in more detail. In addition, at $z=2$, the
observed BH mass function is peaking at $\log(M_\bullet/M_\odot) \sim
8$ and has a number density of $\log(\Phi) \sim -3$ for
$\log(M_\bullet/M_\odot) = 7$ and a number density of $\log(\Phi) \sim
-5$  for $\log(M_\bullet/M_\odot) = 9$. This is also predicted by
the 68Mpc/uhr run.}

\subsection{The BH-bulge mass relation}\label{BH-bulge}

The top panel of Fig. \ref{BHbulgerel} illustrates another fundamental
property of BHs, the BH-stellar mass relation in the present-day
Universe  (colour-coded 2-dimensional histogram, the mean
relation is indicated by the red line). For simplicity, we have used
the total stellar mass and not only the bulge mass, which is
considered in observations. However, in our simulations, galaxies with
masses above $\log(M_{\mathrm{stellar}}/ M_{\odot}) >  10.5$ are largely
systems only consisting of a spherical component as the resolution is
not high enough to provide sufficient morphological information of the 
galaxies. Compared to observations from \citet{Haering04}  (black
lines and symbols), we find an excellent agreement with the
simulations. {This is a direct consequence of the choice of the feedback
efficiency, which in counter-play to the cooling, sets the self regulated
state of the late time evolution (see \citealp{Churazov05}).} 

{In the bottom panel of Fig. \ref{BHbulgerel} we show the evolution of
the BH-stellar mass relation at $z=0-4$ for the 500Mpc/hr (coloured
solid lines) and the 68Mpc/uhr run (coloured dashed lines). We find no
significant effect of resolution on the relation and thus, in both
runs, the relation is in place at $z=3$ (see turquoise lines).} Below
$z=3$, the slope of the relation is similar to the one of the present-day
relation, but the BH masses are mainly found to be under-massive for
a given stellar mass. 

Between $z=1$ and $z=3$, the BH-to-stellar mass ratio is
slightly increasing for a given stellar mass, i.e. BHs are
growing slightly faster than the corresponding galaxy stellar
masses. In contrast, afterwards ($z<1$), this ratio decreases until
today, i.e. the relation is just shifted towards more massive galaxy
masses, what is most likely due to the strong late growth of massive
galaxies. Overall, this is mainly consistent with previous results from
cosmological simulation (\citealp{DiMatteo08}).

\subsection{The SFR and BH accretion rate densities}\label{Madau}

Observations reveal that star formation rate (SFR) and BH accretion
rate densities ($\dot{\rho}_{\mathrm{stellar}}$ and $\dot{\rho}_{\bullet}$, 
respectively) trace each other over cosmic time with
$\dot{\rho}_{\mathrm{stellar}} \sim 2 \times 10^3 \times 
\dot{\rho}_{\bullet}$ (e.g. \citealp{Zheng09} and references 
therein). {In Fig. \ref{madauplot} we show the predicted cosmic
  evolution of the SFR (orange lines) and BH accretion rate densities
  (red lines, multiplied with a factor of $2 \times 10^3$) for the
  500Mpc/hr (solid lines) and for the 68Mpc/uhr run (dashed lines).
At high redshifts $z>1.5$, both the SFR and BH accretion rate 
densities are larger in the 68Mpc/uhr run than in the 500Mpc/hr one.
This explains the increased low mass end of the stellar and BH mass
function at high redshifts in the 68Mpc/hr run (see Figs. \ref{Evolstellar}
and \ref{Evolbh}), where star formation and gas accretion onto
BHs is more efficient.}

In both runs, SFR and BH accretion rates densities peak between
$z = 1-2$  followed by a decline below $0.1\ M_\odot\ \mathrm{yr}^{-1}\
\mathrm{Mpc}^{-3}$ after $z=1$ (for the 500Mpc/hr run). This is
qualitatively consistent with the observational compilation for the
star formation rate densities derived from different wavebands
(coloured squares: \citealp{HopkinsA06}). At high redshifts $z>4$,
however, simulated BH accretion rate densities are too low  compared
to the observed star formation rate densities, which is a consequence
of resolution (see improvement for the 68Mpc/uhr run shown by the
dashed red lines).   
Between $z=2-4$, both the BH accretion and SFR are slightly higher
compared to the observational data, particularly in the higher
resolution run. Below $z=1$, the BH accretion rate densities are in
good agreement with the observed data (for the SFR), whereas the
simulated SFR densities are too high (by a factor of 3-4 at $z=0$) --
a consequence of the too inefficient radio-mode feedback
implementation resulting in too massive present-day galaxies, as
discussed earlier in Fig. \ref{Evolstellar}.  

\begin{figure*}
\begin{center}
  \epsfig{file=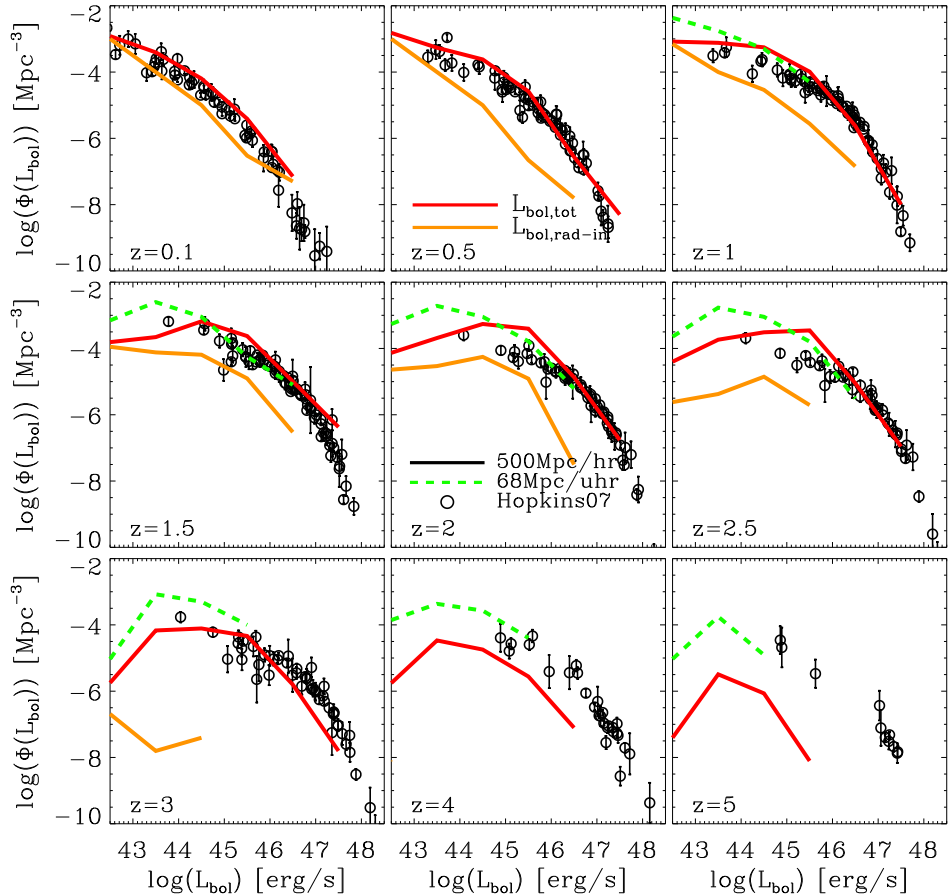,
    width=0.75\textwidth} 
  \caption{Evolution of the bolometric AGN luminosity function for the
    500Mpc/hr (red solid lines) and the 68Mpc/uhr run (green dashed
    lines) at $z=0-5$. Simulation predictions of the 500Mpc/hr run
    match the observational data of \citet{Hopkins07} (black, open
    circles) reasonably well until $z=3$, even if at $z=1.5-2.5$ the
    low luminosity end is over-estimated by up to one dex. At higher z
    $z=3-4$,  the amount of AGN is under-estimated in the 500Mpc/hr
    run due to resolution effects, while the amount of faint AGN in the
    68Mpc/uhr run is increased by 1~dex providing a better match to
    the observational compilation. The orange lines illustrate
      the increasing contribution to the total luminosity by
      radiatively inefficient AGN with decreasing redshift.} {\label{QLF_lowz}} 
\end{center}
\end{figure*}
\begin{figure*}
\begin{center}
  \epsfig{file=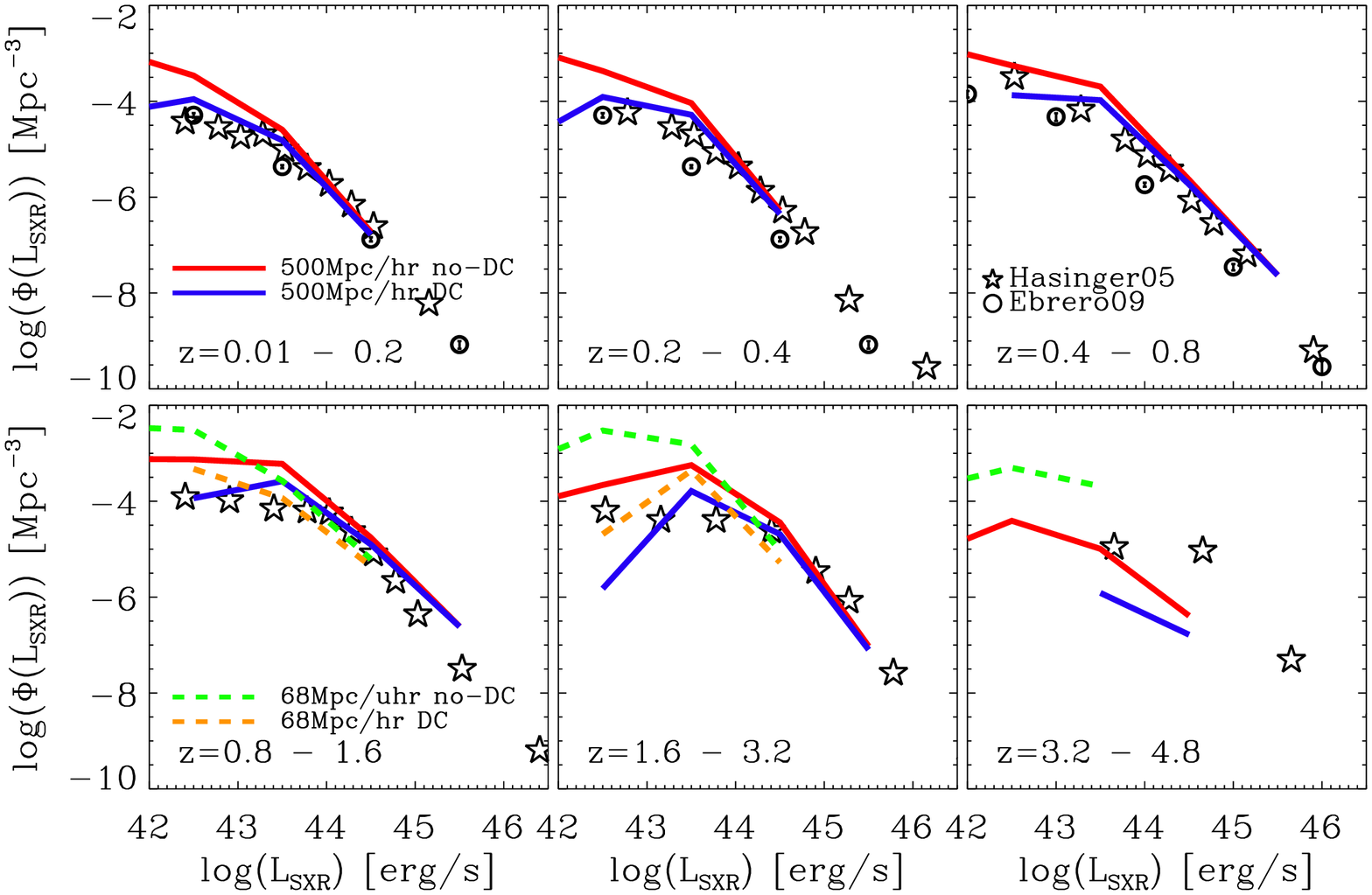,
    width=0.8\textwidth} 
  \epsfig{file=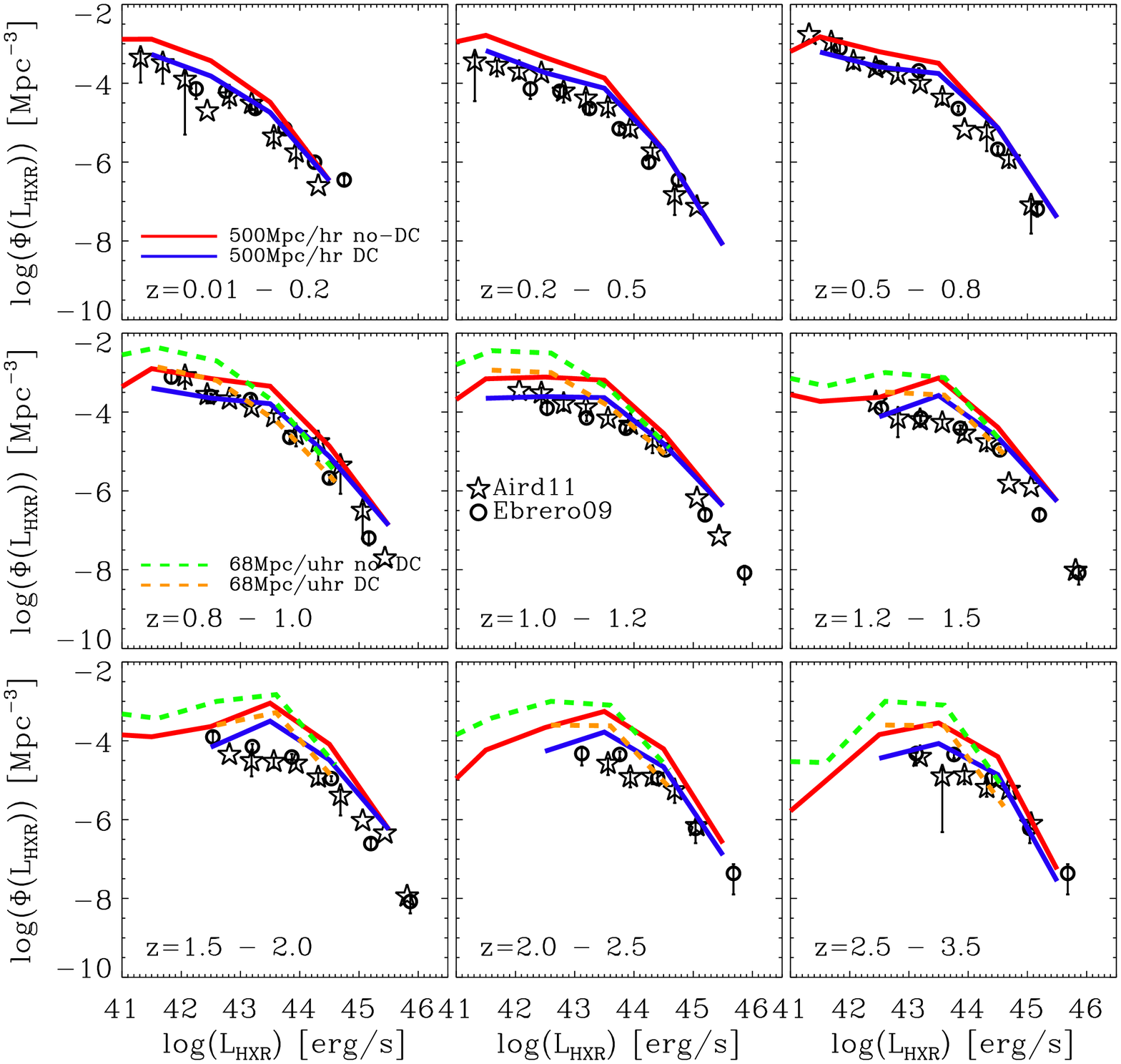,
    width=0.8\textwidth} 
  \caption{First and second rows: Evolution of the soft X-ray
    ($0.5-2$~keV) AGN luminosity function at $z=0-4.8$. Third, fourth
    and fifth rows: Evolution of the hard X-ray ($2-10$~keV) AGN
    luminosity function at $z=0-3.5$. Blue (orange) and red (gree)
    solid (dashed) lines illustrate the results of the 500Mpc/hr
    (68Mpc/uhr) run with and without accounting for dust obscuration,
    respectively. For soft and hard X-ray luminosities at $z=0-1.6$
    and at $z=0-3.5$, respectively, we find a good match with
    observational data (black symbols, \citealp{Hasinger05, Ebrero09,
      Aird10}) when accounting for dust obscuration.}   
  {\label{SHXR_lowz}}  
\end{center}
\end{figure*}

\section{The AGN luminosity function}\label{AGNlum_lowz} 

In this section, {we focus on both radiatively efficient and
inefficient AGN in the 500Mpc/hr and the 68Mpc/uhr run} by
comparing their bolometric luminosities at redshifts between $z=0-5$ 
as well as their present-day radio luminosities to observational
data. For an improved comparison with observations, we also derive the
soft and hard X-ray luminosities and adopt a dust obscuration model
(on a torus level) being dependent on both redshift and AGN
luminosity.     

\subsection{Bolometric luminosities}\label{bollum}

{The bolometric luminosity $L_{\mathrm{bol}}$ (=radiative
  luminosity) is calculated following the study of
  \citet{Churazov05}, who split the total energy released by an AGN
  into to a radiative and mechanical (outflow) component. The
  radiative component is high for radiatively efficient AGN with
  Eddington-ratios $f_{\mathrm{Edd}} = \dot{M}_\bullet /
  \dot{M}_{\bullet,\mathrm{Edd}} > 0.1$, while it is decreasing for
  radiatively inefficient AGN with $f_{\mathrm{Edd}} <0.1$. For
  radiatively efficient AGN, the bolometric luminosity is calculated
  according to
\begin{equation}
  L_{\mathrm{bol}} = \frac{\epsilon_r}{1-\epsilon_r} \dot{M}_\bullet c^2,
\end{equation}
 while for radiatively inefficient AGN the bolometric luminosity is 
  computed by (following Fig. 1 in \citealp{Churazov05}):
\begin{equation}\label{Lineff}
L_{\mathrm{bol}} = 0.1 \times L_{\mathrm{Edd}} \times (\dot{M}_\bullet/
\dot{M}_{\bullet,\mathrm{Edd}} \times 10)^2.
\end{equation}
This distinction is often neglected when calculating the bolometric
luminosity for AGN in simulations and/or semi-analytic models in
literature, where the bolometric luminosity is typically always
computed as for the radiatively efficient case
(i.e. Eq. \ref{Lrad}). The modification used in this work slightly
lowers the amount of faint AGN.}

Fig. \ref{QLF_lowz} shows the bolometric luminosity function at
different redshift steps $z=0.1-5$ in the 500Mpc/hr (red solid lines)
and the 68Mpc/uhr run (green dashed lines). Our simulation results
spanning a broad luminosity range of $10^{42.5} \mathrm{erg}\
\mathrm{s}^{-1} < L_{\mathrm{bol}} < 10^{48} \mathrm{erg}\
\mathrm{s}^{-1}$ are compared to the observational compilation of
\citet{Hopkins07} (black circles). In this study, they convert the AGN
luminosities from different observational data sets and thus, from
different wavebands (emission lines, NIR, optical, soft and hard
X-ray) into bolometric ones. 

{Up to $z=3$, we obtain a reasonably good agreement between the 
500Mpc/hr simulation run and the observations, particularly for the
high-luminous end. At $z=1.5-2.5$, however, AGN with luminosities 
below $L_{\mathrm{bol}} < 10^{45} \mathrm{erg}\ \mathrm{s}^{-1}$ are
slightly over-estimated, a trend which is worse in the 68Mpc/uhr run
due to increased resolution. In contrast, for higher luminosities,
($L_{\mathrm{bol}} > 10^{45} \mathrm{erg}\ \mathrm{s}^{-1}$) the
predictions of the 68Mpc/uhr run are in agreement with the lower
resolution simulation. This suggests that -- at least for the luminous
end of the luminosity function -- the simulation predictions with
increasing resolution seem to {converge against} the observational 
data.} 

The contribution to the total bolometric luminosity by
  radiatively inefficient AGN (their luminosity is calculated
  according to equation \ref{Lineff}, illustrated by the orange, solid
  lines) is significantly increasing down to $z=0$ and is particularly
  dominating the low-luminosity end at lower redshifts. This behaviour
  is mainly driven by an increasing amount of massive BHs accreting
  lower Eddington-ratios as we will discuss in section
  \ref{Connectmbhlum} in more detail. 

{Above $z = 3$, the 500Mpc/hr simulation run starts to
underestimate the amount of AGN for the entire luminosity range by up
to two orders of magnitude (at $z=5$). This can be largely seen as a
consequence of a combination of both insufficient resolution for
accretion in mainly low-mass BHs and too massive seed BHs (only a
minor issue). Instead, for the 68Mpc/uhr run (green dashed lines),
the amount of low luminous AGN is significantly increased resulting in
a reasonably good agreement with the faintest AGN in the observations
at $z=4$. }

{At $z=5$, in order to provide a fair comparison between observations
and simulations, a larger cosmological volume would be needed combined
with an increased resolution than currently adopted in the 500Mpc/hr
and the 68Mpc/uhr runs. For the present, this, however, represents a
great challenge for the currently available computational power. In a
recent study of \citet{Degraf12}, they have performed cosmological
simulation (using Gadget2 with a similar model for BH growth) with a
large box-size of $(500\ \mathrm{Mpc}\ h^{-1})^3$, but only run down
to $z=5$. When considering the AGN  luminosity function at $z=5$ and
$z=6$ they obtain a fairly good agreement to observational data at
these high redshifts.  } 

Nevertheless, the overall good agreement between simulations and
observations (up to $z=3-4$) may indicate that the BH growth closely
follows the gas density and other physical quantities included in the
Bondi-Hoyle accretion formula Eq. \ref{Bondi} in the resolved
vicinity of the BH. This seems to be the case, although simulations
are not able to capture the physical processes on small scales,
i.e. they do not resolve the inner parts of the galaxy
  (3.75~kpc$/$h) and thus, adopt a very rough approximation for the
accretion process itself. We will discuss this in more detail in
section \ref{Origin}.    

A previous study of \citet{Degraf10} also investigated the evolution
of the AGN luminosity function in cosmological simulations with
similar and larger resolution to ours. However, their simulations have
a significantly smaller box-size so that they  can only probe the
low-luminosity end of the AGN luminosity function (up to
$L_{\mathrm{bol}} < 10^{45}\ \mathrm{erg}\ \mathrm{s}^{-1}$), which
they find to be in reasonably good agreement with
observations. {In a very recent work of \citet{Khandai14} they
  analyse a simulation with a larger boxsize of $(100\ \mathrm{Mpc}\
  h^{-1})^3$. They can match the low-luminosity end
  ($L_{\mathrm{bol}} < 10^{45}\ \mathrm{erg}\ \mathrm{s}^{-1}$) up to
  $z=4$, but the luminous end is either significantly over-estimated
  (at $z<0.5$) or under-estimated (at $z>0.5$).}

\subsection{X-ray luminosities}\label{Xlum} 

By now, we compared our simulation predictions to the observational
compilation of \citet{Hopkins07} (see Fig. \ref{QLF_lowz}).
{In this study, they assume a \textit{luminosity} dependence of
the obscured fraction (the less luminous the more obscured) and the
same number of Compton-thick ($N_H > 10^{24}\ \mathrm{cm}^{-2}$) and
Compton-thin ($10^{23}\ \mathrm{cm}^{-2}< N_H < 10^{24}\
\mathrm{cm}^{-2}$) AGN. However, there are many aspects of the
obscuration corrections that are still being vigorously debated. }

{Some recent studies suggest that the obscured fraction is dependent on
both luminosity and redshift (\citealp{Hasinger08, Fiore12}), in
contrast with the non-redshift dependent model of
\citet{Hopkins07}. There are also uncertainties with respect to the
dust correction for the UV luminosity; \citet{Hopkins07} compute the
amount of dust (and therefore extinction), by adopting an $N_H$
distribution from X-ray observations, and a Galactic dust-to-gas
ratio. However, it has been shown that AGN absorbers do not have a
Galactic dust to gas ratio  (\citealp{Maiolino01, Maiolino04}). The
result is that they probably over-estimate the extinction, which might
result in slightly higher luminosities for the optically selected
quasars. Because of these uncertainties, we additionally attempt in
this subsection to correct our model predictions for obscuration and
to directly compare them with recent soft ($0.5-2$~keV) and hard
($2-10$~keV) X-ray measurements of the AGN luminosity function
(\citealp{Hasinger05, Ebrero09, Aird10}).}

{In contrast to the previous subsection, we do not attempt to
  correct the observations for obscuration (as it is done for the
  observational compilation in \citealp{Hopkins07}), but instead apply
  an obscuration correction to our models.} We convert the modelled,
bolometric luminosities into hard and soft X-ray luminosities ($0.5-2$
keV and $>2$ keV) using the bolometric correction according to
\citet{Marconi04}. In their study, the hard and soft X-ray
luminosities  $L_{\mathrm{HXR}}, L_{\mathrm{SXR}}$ are approximated by
the following third-degree polynomial fits: 
\begin{eqnarray}
\log(L_{\mathrm{HXR}}/L_{\mathrm{bol}}) = -1.54 - 0.24\mathcal{L} -
0.012 \mathcal{L}^2 + 0.0015 \mathcal{L}^3\\
\log(L_{\mathrm{SXR}}/L_{\mathrm{bol}}) = -1.65 - 0.22\mathcal{L} -
0.012 \mathcal{L}^2 + 0.0015 \mathcal{L}^3
\end{eqnarray}
with $\mathcal{L} = \log(L_{\mathrm{bol}}/L_{\odot}) - 12$.  These
corrections are derived from template spectra, which are truncated at
$\lambda > 1\ \mu m$ in order to remove the IR bump and which are
assumed to be independent of redshift (therefore the resulting
bolometric corrections are also assumed to be redshift independent).

Additionally, we apply a correction for obscuration to the model
luminosities, as suggested by several observational studies
(\citealp{Ueda03, Hasinger04, LaFranca05}), in which it has been shown
that the fraction of obscured AGN is luminosity dependent and
decreases with increasing luminosity. While older studies such as
\citet{Ueda03} and \citet{Steffen03} did not find a clear dependence
of obscuration on redshift, several recent observational studies
(\citealp{Ballantyne06, Gilli07, Hasinger08}) propose a strong
evolution of the obscured AGN population (with the relative fraction
of obscured AGN increasing with increasing redshift). Here, we follow
the study of \citet{Hasinger08}, where they compare the same AGN in
both the soft and hard X-ray band so that they can derive an
approximation for the obscured fraction $f_{\mathrm{obsc}}$ in the
soft X-ray band. 

The obscured fraction at $z<2$ is then given by this  equation: 
\begin{eqnarray}\label{obsc1}
f_{\mathrm{obsc}} (z,L_{\mathrm{SXR}}) = -0.281
(\log(L_{\mathrm{SXR}})-43.5) + \nonumber \\
+ 0.279(1+z)^{\alpha},
\end{eqnarray}
where $L_{\mathrm{SXR}}$ is the soft X-ray luminosity and they find
that a value of $\alpha = 0.62$ provides the best fit to their
observational data. The obscured fraction at $z>2$ is approximately
the same at $z=2$:  
\begin{eqnarray}\label{obsc2}
f_{\mathrm{obsc}} (z,L_{\mathrm{SXR}}) = -0.281
(\log(L_{\mathrm{SXR}}) - 43.5) + 0.551.  
\end{eqnarray}
Note that according to equations \ref{obsc1} and \ref{obsc2}, the obscured
fractions can get negative or larger than 1, why we additionally
impose that negative values are equal zero and we set values larger
than 1 equal to 1. By calculating the obscured fraction of AGN in the
soft X-ray band, we can model the visible fraction of AGN
$f_{\mathrm{vis}} = 1 - f_{\mathrm{obsc}}$ and thus, the visible
number density of AGN in the soft X-ray range is given by:  
\begin{eqnarray}
\Phi_{\mathrm{vis}}(L_{\mathrm{SXR}}) = f_{\mathrm{vis}} \times
\Phi_{\mathrm{total}}(L_{\mathrm{SXR}}) 
\end{eqnarray}

\begin{figure}
\begin{center}
  \epsfig{file=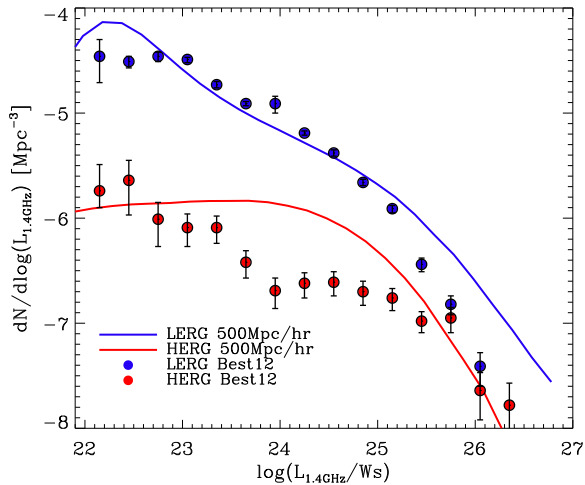,
    width=0.45\textwidth} 
  \caption{Radio luminosity function at $z=0$ distinguishing between
    high excitation radio galaxies (red line) and low excitation radio
  galaxies (blue line) in the 500Mpc/hr simulation. Simulation
  predictions are widely in agreement with observational data of
  \citet{Best12} (coloured symbols), for both high and low excitation
  radio galaxies implying that the simulations predict the correct
  amount of radiatively efficient and inefficient radio AGN.}
{\label{Radio_lum}} 
\end{center}
\end{figure}

{The first and second rows in Fig. \ref{SHXR_lowz} show the soft
  X-ray luminosity function at $z=0.01-4.8$ (different panels) in the
500Mpc/hr (solid lines) and 68Mpc/uhr (dashed lines) simulation runs
without (red and green lines) and with dust obscuration (blue and
orange lines) and in observations (black stars and circles:
\citealp{Hasinger05} and \citealp{Ebrero09}). At the high-luminosity end
($L_{\mathrm{SXR}} > 10^{44}\ \mathrm{erg}\ \mathrm{s}^{-1}$),
obscuration  does not influence the results and thus, the same trends
can be seen as already discussed in Fig. \ref{QLF_lowz}, i.e. a good
match with the observations. }

{However, turning to the low-luminosity end, at $z<3.2$, simulation
results without dust correction over-estimate the amount of moderately 
luminous AGN by up to 1~dex in the 500Mpc/hr run and by up to 2~dex in
the 68Mpc/uhr run. Instead, the simulation predictions including dust
obscuration can achieve a fairly good agreement with the
observations. This implies that our results are consistent with the
observational conclusions of \citet{Hasinger08} supporting the
adoption of a redshift-dependent dust obscuration. }

Note that the adopted dust obscuration model is the same as the one in
a recent study of \citet{Hirschmann12} and similar to the one adopted
in \citet{Fanidakis12}, where they use the semi-analytic model of
\citet{Somerville08} and \citet{Bower06}, respectively, to study the
evolution of the AGN luminosity function. In these works, they also
find a good agreement between models and observations when adopting a
redshift \textit{and} luminosity dependent dust obscuration and
comparing their model predictions to observed soft X-ray luminosities.   

{The third to fifth rows of Fig. \ref{SHXR_lowz} illustrate the
\textit{hard} X-ray luminosity functions predicted by the two
simulation runs for different redshift ranges $z=0.01-3.5$, compared
with observational data (\citealp{Ebrero09}, \citealp{Aird10}). The
simulation predictions without dust obscuration (red solid and green
dashed lines) are able to reproduce the high-luminosity end pretty
well, while the amount of fainter AGN is strongly over-estimated,
particularly in the 68Mpc/uhr run above $z=1$.}

One possible explanation might be again due to obscuration as even
$2-10\ \mathrm{keV}$ X-ray surveys might miss a significant fraction
of moderately obscured AGN ($~ 25$\% at $N_H = 10^{23}\
\mathrm{cm}^{-2}$) and nearly all Compton-thick AGN ($N_H > 10^{24}\
\mathrm{cm}^{-2}$, \citealp{Treister04, Ballantyne06}). From fits to
the cosmic X-ray background, \citet{Gilli07} predict that both
moderately obscured and Compton-thick AGN are as numerous as
unobscured AGN at luminosities higher than $\log(L_{0.5-2{\rm keV}})>
43.5\ [\mathrm{ergs/s}]$, and four times as numerous as unobscured AGN
at lower luminosities ($\log(L_{0.5-2 {\rm keV}}) < 43.5\
[\mathrm{ergs/s}]$).  

In addition, in a study of \citet{Merloni14}, they investigate a
complete sample of 1310 AGN from the XMM-COSMOS survey between
$z=0.3-3.5$ and classify them to be obscured in the hard X-ray band
due to the shape of the X-ray spectrum\footnote{objects are 
  selected to be obscured with a column density larger than
  $N_H>10^{21.5}\ \mathrm{cm}^{-2}$}. Their results confirm earlier
results that at each redshift there is a clear decrease of the
fraction of obscured AGN with increasing luminosity. In addition (and
in contrast to an optical classification), nuclear obscuration in the
hard X-ray band also reveals a dependence on redshift, where the
fraction of obscured AGN increases with increasing redshift. 

For this reason, we considered the corresponding observed, (hard
X-ray) obscured fractions of AGN at a given redshift and luminosity
(table 1 in \citealp{Merloni14}) in our simulations. Due to the
limited redshift range, we have assumed that obscured fractions at
$z<0.3$ are the same as at $z=0.3$ and that the obscured fractions at
$z>3.5$ stay also constant. This results in a fairly good match to the
observed low-luminosity end at $z=0.01-1.5$ (blue solid and orange
dashed lines in Fig. \ref{SHXR_lowz}). At higher redshifts $z>1.5$,
moderately luminous AGN are still somewhat too numerous in the
simulations (even if improved), particularly in the 68Mpc/uhr
run. Nevertheless, at $z\sim0-3.5$, our simulation results are widely
consistent with the existence of a redshift and luminosity dependent
obscured fraction of AGN in the hard X-ray band as suggested by
observational data of \citet{Merloni14}.

\subsection{The radio luminosity function at $z=0$}\label{Radiolum} 

In the last subsections, we were focusing on radiative AGN
luminosities above $10^{43}$~erg/s with the main contribution from
radiatively efficient AGN. Radiatively less efficient AGN, however,
are often observed to produce energetic, powerful radio jets and as a
consequence hot X-ray cavities and thus, heat the interstellar
medium. Therefore, in this section we explicitly investigate the
present-day radio luminosity function by distinguishing between low
excitation (``radio-mode'', LERG) and high excitation
(``quasar-mode'', HERG) radio galaxies with Eddington-ratios
$f_{\mathrm{edd}} < 0.01$ and $f_{\mathrm{edd}} > 0.01$,
respectively. This allows us to study particularly all the AGN
accreting at very low Eddington-ratios and to quantify their number
density.  

To derive a radio luminosity, we follow a recent observational study
of \citet{Best12} who investigate a sample of $18,286$ radio-loud AGN
by combining data from different  surveys and derive the radio
luminosity functions for the first time separately for the LERGs and
HERGs. In their Fig. 6, they show histograms for the ``total''
Eddington-ratios $f_{\mathrm{Edd,tot}} = (L_{\mathrm{rad}} +
L_{\mathrm{mech}}) / L_{\mathrm{edd}}$ of the HERGs and the LERGs, 
which are peaking at ratios of $\sim -1.6$ and $\sim -3$,
respectively. {We adopted Gaussian distributions to model
  these ``total'' Eddington-ratio distributions for HERGs and  LERGs with a
  standard deviation of 0.3 and 0.4, respectively. For each AGN, we
  randomly draw a ``total'' Eddington-ratio so that we can derive the
  mechanical jet luminosity $L_{\mathrm{mech}}$ for the HERGs and the
  LERGs:} 
\begin{eqnarray}
L_{\mathrm{mech,HERGs}} & = & f_{\mathrm{Edd,tot,HERGs}} \times L_{\mathrm{edd}} -
L_{\mathrm{rad}}\\
L_{\mathrm{mech,LERGs}} & = & f_{\mathrm{Edd,tot,LERGs}} \times L_{\mathrm{edd}} -
L_{\mathrm{rad}}
\end{eqnarray}
$L_{\mathrm{edd}}$ is the maximum Eddington luminosity and
$L_{\mathrm{rad}}$ the radiative (bolometric) luminosity of an AGN.
Using the relation of \citet{Cavagnolo10} between the $1.4$~GHz radio
luminosity $L_{1.4\mathrm{GHz}}$ and the mechanical AGN jet
luminosity, which is given by  
\begin{equation}
L_{\mathrm{mech,HERGs/LERGs}} = 10^{36} \times (L_{1.4\mathrm{GHz}} /
10^{24} \mathrm{WHz}^{-1})^{0.7} \mathrm{W},
\end{equation}
we can solve for the observed 1.4~GHz radio luminosity for both the
HERGs and the LERGs. 

Fig. \ref{Radio_lum} shows the present-day radio luminosity function
of the 500Mpc/hr simulation (solid lines) and of observational data of
\citet{Best12} (coloured circles) distinguishing between LERGs (blue)
and HERGs (red). In both observations and simulations, the LERGs
provide the dominating contribution to the overall radio
luminosity. We find a reasonably good agreement between simulations
and observations for both samples of radio galaxies. This implies that 
simulations are also able to predict reasonable radio luminosities and
moreover, to reproduce the correct amount of radiatively efficient and
inefficient radio AGN. This agreement also indirectly suggests that
the threshold value for the accretion rate separating quasars and
radio mode is a sensible one.

\begin{figure}
\begin{center}
  \epsfig{file=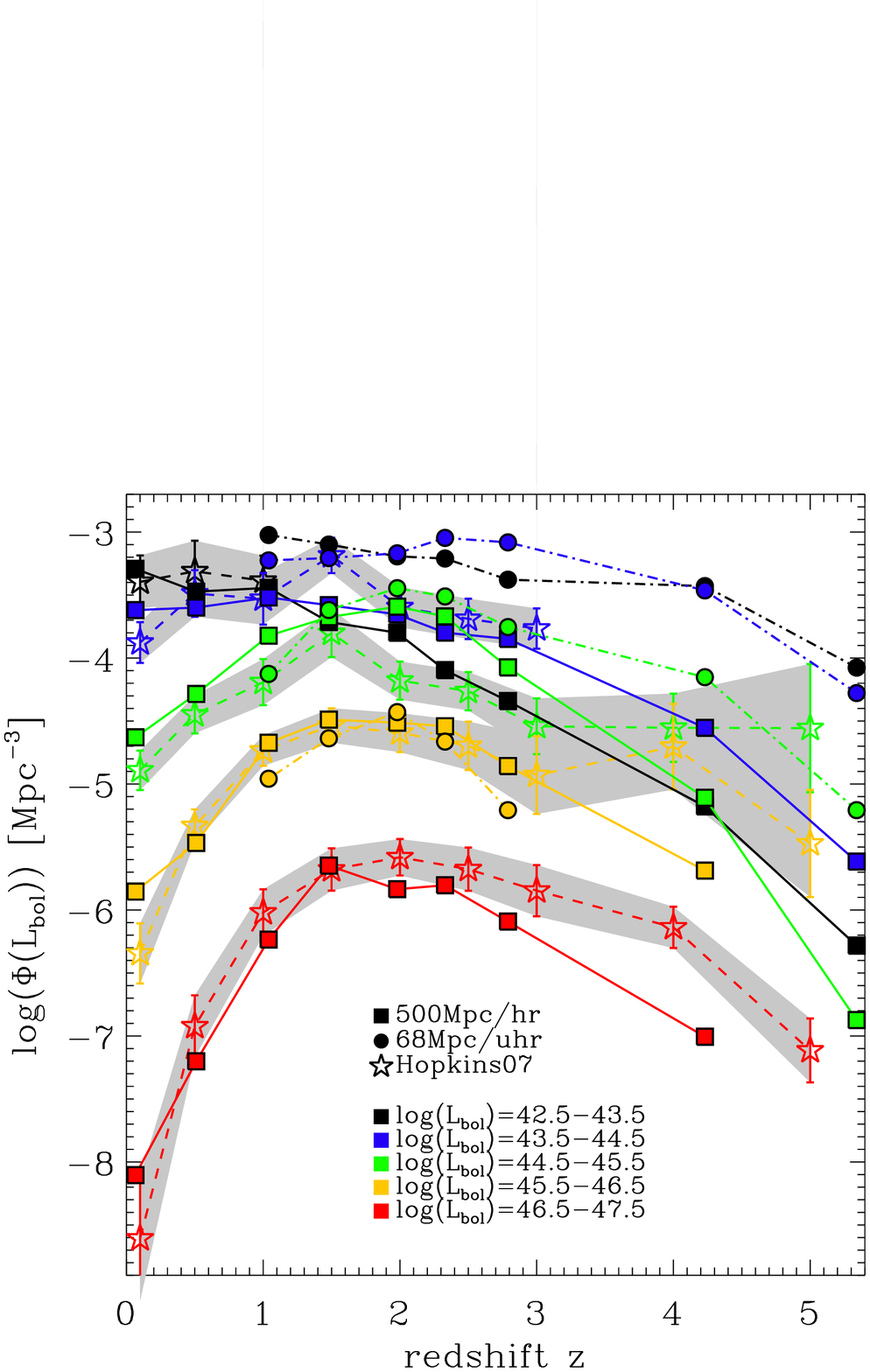,
    width=0.5\textwidth} 
  \caption{Evolution of the co-moving number density of AGN binned in    
    different bolometric luminosity bins (indicated by different
    colours) in the two simulation runs (filled squares and solid
    lines: 500Mpc/hr; filled circles and dashed-dotted lines:
    68Mpc/uhr) and in the observational compilation of
    \citet{Hopkins07} (open stars with dashed lines and the grey
    shaded areas). The Simulations successfully predict the observed
    anti-hierarchical trend, i.e. the number density of luminous
    AGN peaking at higher redshifts than the one of less luminous ones.} 
  {\label{Num_dens}} 
\end{center}
\end{figure}

\section{Anti-hierarchical growth of BHs}\label{Anti-hier} 

\subsection{Co-moving number density evolution of AGN}\label{Numdens} 

In the sections \ref{AGNlum_lowz} we have shown that a combination of
the 500Mpc/hr and the 68Mpc/uhr simulation runs can reproduce the
evolution of the AGN luminosity function from $z=4$ to $z=0$
reasonably well. This implies that our simulations should also be able
to naturally reproduce the observed ``anti-hierarchical'' or
``downsizing'' trend in BH growth within the framework of a
hierarchical structure formation scenario. As described in the
introduction, the term ``downsizing'' refers to the observational
result that the number density of luminous AGN peaks before
successively less luminous AGN. 

This is visualised in Fig. \ref{Num_dens} showing the redshift
evolution of the AGN number densities binned in different bolometric
luminosity bins (different coloured lines as indicated in the
legend). The observational data of \citet{Hopkins07} (open stars with
the dashed lines and the grey shaded area) reveal a time evolution of
the peaks of the luminosity curves being characteristic for the
downsizing trend. At $z<3$, the 500Mpc/hr run (filled squares with
solid lines) matches the observations reasonably well (only slightly
over-predicts the amount of moderately luminous AGN), while at higher
redshifts, the 68Mpc/uhr run (filled circles with dotted-dashed lines)
agrees better with the observations (due to higher resolution), but
only for moderately luminous AGN (due to small box size). 

This clearly shows that the simulations can correctly capture the
observed time evolution of the peaks of the different luminosity
curves, and thus, the corresponding downsizing trend. The strong
decrease of luminous AGN (see yellow and red curves) at low redshift
$z<2$ is mainly due to the decreasing cold gas content in the vicinity
of the massive BHs, i.e. if a merger happens (which is typically
supposed to induce nuclear activity), it is preferentially a 'dry'
merger not triggering any significant AGN activity. Instead,
moderately luminous AGN probably are either low-mass BHs with high
accretion rates (close to the Eddington-rate) or massive BHs, which
are - due to the small amount of cold gas which is left - accreting at
low rates and which have most likely been very luminous in the
past. This will be discussed and analysed in more detail in the next
sections.  

\begin{figure*}
\begin{center}
  \epsfig{file=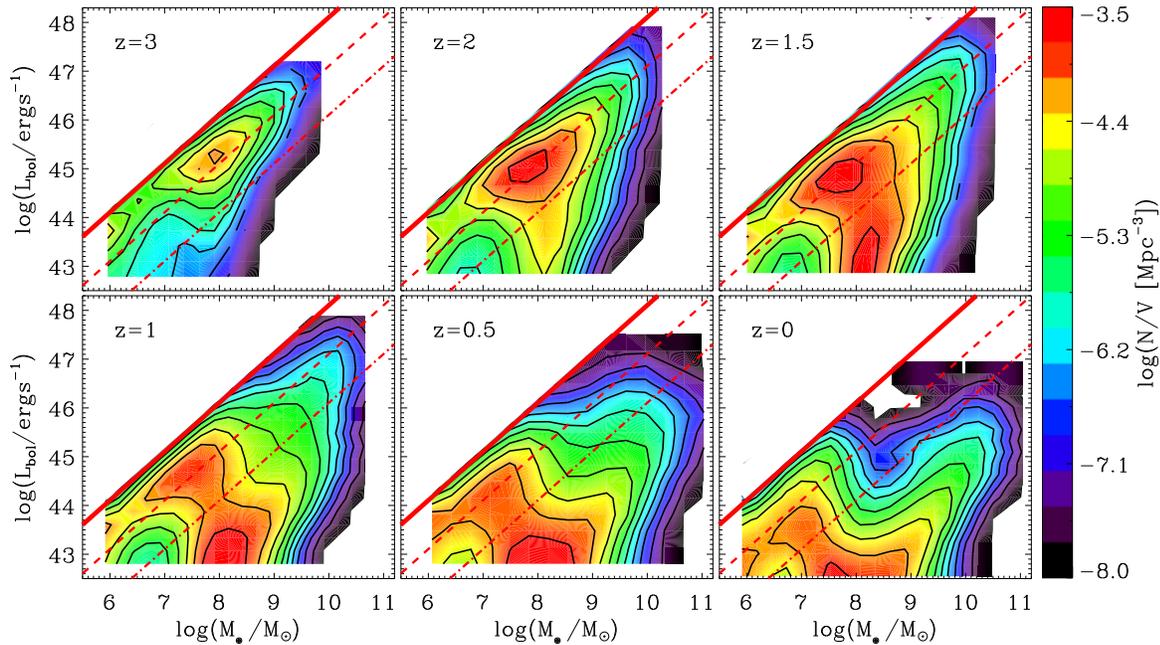, width=0.9\textwidth}
  \caption{2-dimensional histograms (number density is colour-coded) of
    the bolometric luminosity-BH mass-plane at different
    redshifts $z=3-0$ for the 500Mpc/hr simulation run. To guide the
    eye, the red lines indicate an accretion at the Eddington rate
    (solid line), at $1/10$ (dashed line) and at $1/100$
    (dotted-dashed line) of the Eddington rate. At high redshifts, 
    BHs accreting at or close to the Eddington rate 
    dominate. Instead, towards lower redshifts the contribution of
    massive BHs accreting at very low Eddington-ratios
    significantly increases explaining the decline of the number
    density of luminous AGN with decreasing redshift (see red and
    yellow lines in Fig. \ref{Num_dens}). } 
          {\label{Lum_bh}}
\end{center}
\end{figure*}

\subsection{Implications on the connection between BH mass and
  AGN luminosity}\label{Connectmbhlum}    

In this section, we examine consequences and implications of the
downsizing trend in the AGN evolution on the connection between masses
of active BHs, the Eddington ratios and the bolometric luminosities of AGN.

\subsubsection{The bolometric luminosity-BH mass plane}
 
Fig. \ref{Lum_bh} shows a 2-dimensional histogram (AGN number density
is colour-coded) of the bolometric AGN luminosity-BH mass plane at
different redshifts $z=3-0$ for the 500/hr simulation. The red lines
help to visualise an accretion at the  (maximum) Eddington rate (solid
line), at $1/10$ (dashed line) and at $1/100$ (dotted-dashed line) of
the Eddington rate. 

At $z=3$, the majority of the BHs with masses up to $M_{\bullet} \leq
10^9 M_\odot$ is accreting at or at least very close to the Eddington
rate ($0.1 < f_{\mathrm{edd}} < 1$), i.e. for these high redshifts,
the AGN luminosity is approximately linearly related with BH mass.
Turning to lower redshifts ($z \leq 2$), the number density of AGN
with massive BHs ($M_{\bullet} > 10^{9}M_{\odot}$) accreting a lower
Eddington-ratios has increased with respect to $z=3$. Thus, the
relation between BH mass and bolometric luminosity becomes much 
broader, the linear relation found at higher redshifts is broken, and
BHs with masses $M_{\bullet} > 10^8 M_{\odot}$ can now also
power moderately luminous AGN with $L_{\mathrm{bol}} \sim 10^{43}\
\mathrm{erg/s}$ as they are accreting with Eddington ratios below
$f_{\mathrm{edd}} < 0.01$. 

At $z<2$, the probability for BHs with $M_{\bullet} > 3 \times 10^7 M_{\odot}$
to accrete at Eddington ratios below $f_{\mathrm{edd}} = 0.01$ is even
higher than to accrete at larger Eddington ratios. These BHs most
likely are relics from an earlier, more active phase with higher
accretion rates, which undergo at later times a ``blow-out'' or
``fading'' phase, where the gas gets ejected out of the central
regions around the BH due to AGN feedback (and so the gas
density in the vicinity of the BH is reduced, see the discussion in
section \ref{Origin}). 

This process is shown in different studies of e.g. \citet{Hopkins08a}
who perform a detailed analysis of a large set of isolated galaxy
merger simulations. They present typical light curves of AGN triggered
by mergers, with a first phase of high accretion close to the Eddington
rate, followed by a power-law decline in the accretion rate. We expect
that this is broadly traced within our cosmological simulation and
plan to study the detailed light curves in a higher resolution
simulation in a follow-up study. 

Finally, Fig. \ref{Lum_bh} shows that the number density of luminous
AGN ($L_{\mathrm{bol}} > 10^{45}$~erg/s) is significantly smaller
(almost completely suppressed) at $z=0$ than it is at $z=2$. This is
mainly due to the available gas content, which gets reduced
particularly in massive galaxies over cosmic time. This is also
consistent with observational results of e.g. \citet{Steinhardt10} and
\citet{Kelly13}, who find that the probability of massive BHs
accreting close to the Eddington-rate is significantly reduced at low
redshift.        

\subsubsection{The evolution of the Eddington-ratios}

\begin{figure}
\begin{center}
  \epsfig{file=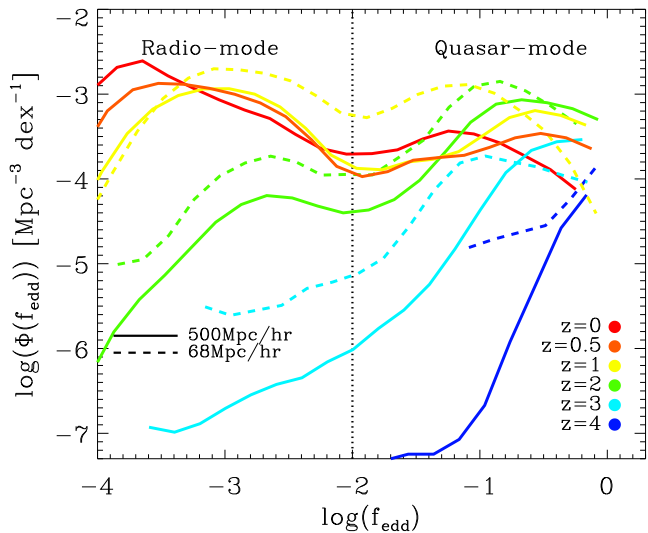, width=0.5\textwidth}
  \caption{Eddington ratio distributions at different redshifts
    $z=0-4$ (coloured lines) for the 500Mpc/hr (solid lines) and the
    68Mpc/uhr simulation run (dashed lines). The vertical dotted line
    visualises the distinction between radio- and quasar-mode adopted
    in the model. The peaks of the distributions are shifted towards
    lower Eddington-ratios with decreasing redshift, in qualitative
    agreement with observational studies (\citealp{Vestergaard03,
      Kollmeier06, Kelly10, Schulze10, Kelly13}).} {\label{hist_fedd}}  
\end{center}
\end{figure}

{Fig. \ref{hist_fedd} shows the redshift evolution of the Eddington
ratio distributions for the 500Mpc/hr (solid, coloured lines) and the
68Mpc/uhr simulation run (dashed coloured lines). The vertical
  dotted line visualises the adopted division between the radio- and
quasar-mode in the simulation code.} Consistent with our previous
analysis, and often adopted luminosity limits of observational
samples, we consider solely AGN with bolometric luminosities larger
than $10^{43}\ \mathrm{erg/s}$. 

The simulated fraction of AGN which are accreting at small
Eddington-ratios ($f_{\mathrm{edd}} < 0.01$) increases strongly with
decreasing redshift and thus, the peaks of the distribution curves are
clearly shifted towards smaller Eddington ratios with decreasing
redshift in both simulation runs -- a typical implication of the
downsizing trend: at $z=4$, the distribution peaks at
$f_{\mathrm{edd}} \sim 1$, while at $z=0$ the peak is located around
$f_{\mathrm{edd}} \sim 5 \times 10^{-4}$. This illustrates the same
trend we discussed for Fig. \ref{Lum_bh}: at later times black holes
spend less time accreting at the Eddington rate, but mainly reside in
a decline dominated ``blowout'' accretion phase.  

As a further consequence, the number of black holes accreting close to
the Eddington-rate is low at $z=0$ ($\log(\mathrm{d}N/
\mathrm{d}f_{\mathrm{edd}}) \sim -4.5\  \mathrm{Mpc}^{-3}\
\mathrm{dex}^{-1}$), while at high redshifts ($z=1-3$) a large number
of BHs ($\log(\mathrm{d}N/ \mathrm{d}f_{\mathrm{edd}})  \sim -3.5\
\mathrm{Mpc}^{-3}\ \mathrm{dex}^{-1}$) are preferentially accreting
within a broad range of Eddington ratios
$0.01<f_{\mathrm{edd}}<1$. To summarise, the majority of AGN at $z=0$
are hardly radiating at or near the Eddington limit, which is in
qualitative agreement with several observational studies
(\citealp{Vestergaard03, Kollmeier06, Kelly10, Schulze10,
  Steinhardt10, Kelly13}). \citet{Kelly10}, for example, show that 
the Eddington ratio distribution (using broad-line quasars between
$z=1-4$) peaks at an Eddington ratio of $f_{\mathrm{edd}} =0.05$.   

\begin{figure}
\begin{center}\vspace{-0.3cm}
  \epsfig{file=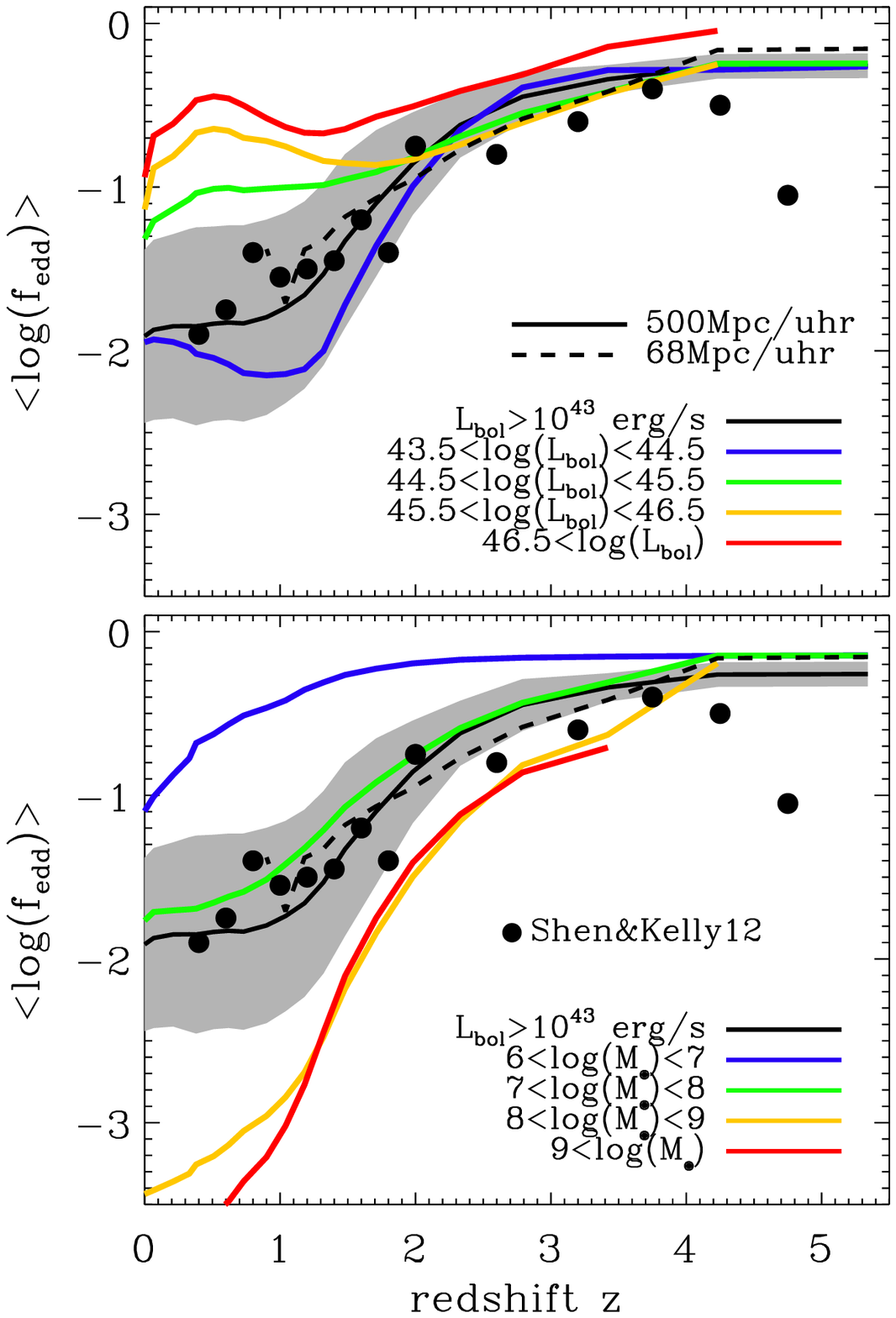, width=0.5\textwidth}\vspace{-0.3cm}
  \caption{Redshift evolution of the mean Eddington-ratios for the
    500Mpc/hr  (black solid lines; grey shaded areas indicate the
    1-$\sigma$ scatter) and the 68Mpc/uhr run (black dashed
    lines). The simulated predictions from both runs match the
    observational mean Eddington-ratios at $z \leq 4$ (black filled
    cirles, \citealp{Shen12}). The 500Mpc/hr
    run is additionally divided into different BH mass bins 
    (coloured lines in the top panel) and into different
    bolometric luminosity bins (coloured lines in the bottom
    panel) showing that luminous AGN and low-mass BHs have on average  
    the highest Eddington-ratios. }
  {\label{fedd_evol}} 
\end{center}
\end{figure}

{In Fig. \ref{fedd_evol} we present a quantitative comparison of
simulated Eddington-ratios in the 500Mpc/hr (solid lines) and the
68Mpc/uhr (dashed lines) run with observational data: the simulated
mean Eddington-ratios of AGN with luminosities above $10^{43}$~erg/s
are plotted versus redshift (see solid and dashed black lines with
grey shaded areas in both panels) and compared to observational data
of \citet{Shen12} (black, filled circles).} In both simulations and
observations, the mean Eddington-ratios are decreasing with decreasing
redshift and we find a nearly perfect agreement between them at $z
\leq 4$. 

Additionally, in the simulations (only explicitly shown for the
500Mpc/hr run), we have divided the AGN sample into different BH mass
bins (solid, coloured lines in the top panel of Fig. \ref{fedd_evol})
and different AGN luminosity bins (coloured lines in the bottom panel
of Fig. \ref{fedd_evol}). This illustrates that on average, for the
entire redshift range since $z \sim 4$, lowest mass BHs (blue
line in the top panel) and most luminous AGN (red line in the bottom
panel) accrete at the highest Eddington-ratios
($0.1<f_{\mathrm{edd}}<1$). In contrast, the most massive BHs  
(red line in the top panel) and the lowest luminous AGN (blue line in
the bottom panel) have the lowest mean Eddington-ratios. This
behaviour is consistent with Fig. \ref{Lum_bh}. 

Finally, we want to point out that the interplay between BH
mass, AGN luminosity and Eddington-ratio, we have discussed in this
section, is in good qualitative agreement with previous studies of
\citet{Fanidakis12} and \citet{Hirschmann12} who also reproduced the
downsizing trend in BH growth, but by using semi-analytic
models. This confirms that such a connection between black hole masses
and AGN luminosities seems to be a necessary condition for reproducing
a downsizing trend in BH growth.   

\subsection{Physical origin of the downsizing trend in our simulations}\label{Origin}    

A common problem of all currently existing, large cosmological
simulations is that they are not able to capture the detailed physical  
processes of gas accretion in the close vicinity of the BH due
to too low resolution as a result of the (up to now) limited
computational power. Besides, the physical description of BH
accretion is also more complicated than typically assumed in
large-scale simulations (e.g. in reality, there is no spherical
symmetry \citealp{King13} and the Bondi-accretion seems to be also a
poor prescription for cold (chaotic) gas accretion,
\citealp{Gaspari13}). Thus, a very rough approximation for the
accretion process (following the Bondi accretion formula) is generally
adopted when performing large-scale simulations as we have described
earlier in section \ref{BHgrowth}.  

Nevertheless, as we have seen in the course of this study, simulations
seem at least to be able to reproduce the right trends of how BHs
assemble their mass and appear as an AGN despite their relatively low
spatial and mass resolution (compared to the physical scales at which
the gas accretion is actually happening). In other words, the general
scheme in our simulations is able to capture the essence of BH
growth ``in reality''. Therefore, to better understand the
\textit{physical reason} for automatically reproducing the downsizing 
trend in our simulations, we directly investigate the different
gaseous quantities, as the density, the sound speed, which is
proportional to the gas temperature and the relative velocity (between
the gas particles and the BH), which are contributing to the
estimation of the Bondi-accretion rate in the code given by
Eq. \ref{Bondi} by averaging over all the gas particles within the
numerically resolved accretion region of the BH. {Note that
``numerically resolved accretion region'' is supposed to describe the
SPH-like smoothing length of the BH sink  particles.}     

\begin{figure*}
\begin{center}
  \epsfig{file=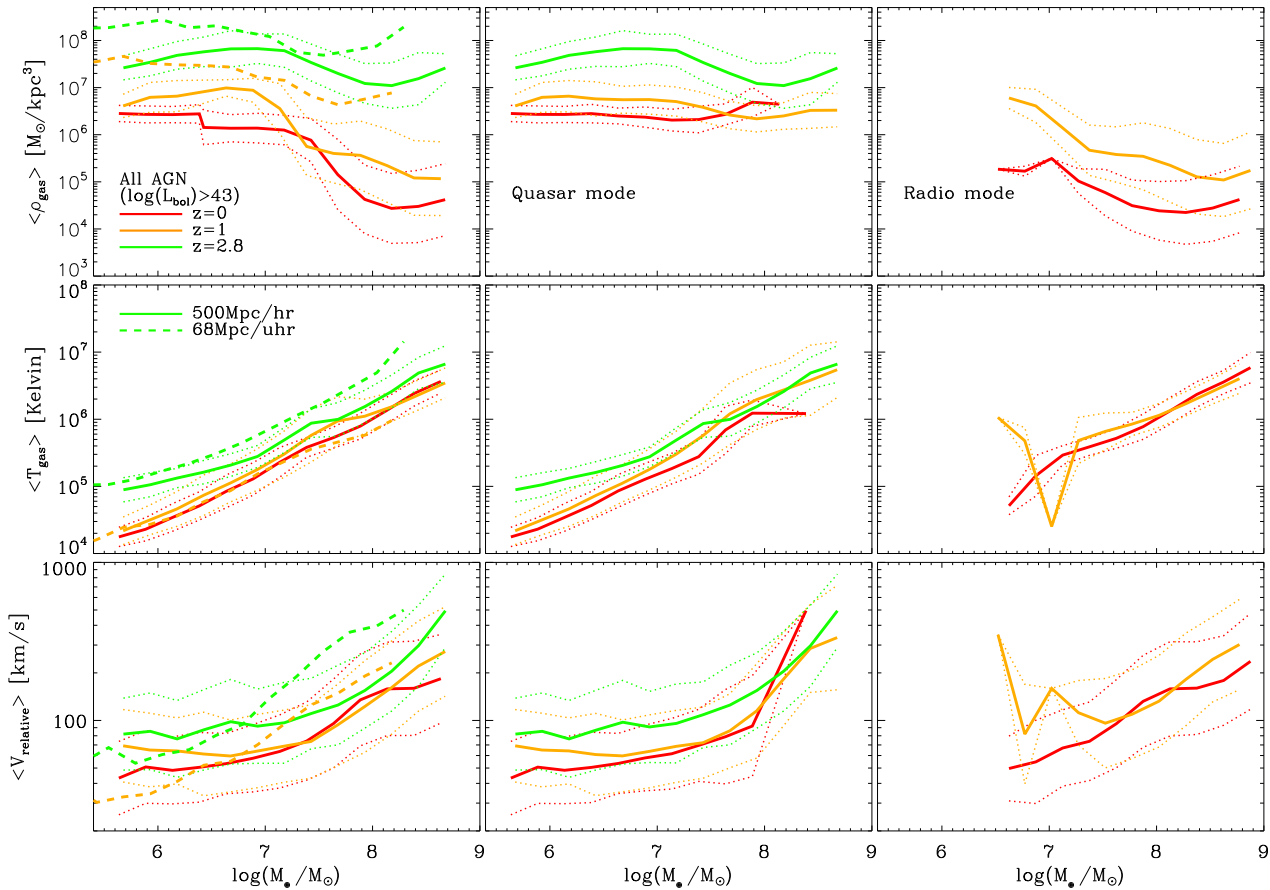, width=0.95\textwidth}
  \caption{Mean gas density (top row), mean gas temperature (middle
    row) and mean relative gas velocity between the BH and the
    gas particles  (bottom row)  within the numerically resolved
    accretion region versus BH mass at $z=2.8$ (green lines),
    $z=1$ (orange lines) and $z=0$ (red lines) for the 500Mpc/hr
    (solid lines) and the 68Mpc/uhr run (dashed lines). The left
    column corresponds to AGN with
    $\log(L_{\mathrm{bol}})>43$~[erg/s], while in the middle and right
    column we additionally distinguish between AGN in the quasar and 
    the radio mode, respectively. Irrespectively of the redshift, the
    gas temperature and relative velocity is increasing, while the gas
    density is decreasing with increasing BH mass. In addition,
  the gas density at a given BH mass is decreasing with decreasing
  redshift.}  
          {\label{Dens_bhmass}}
\end{center}
\end{figure*}

Fig. \ref{Dens_bhmass} shows the mean (solid lines) and the
corresponding 1-$\sigma$  scatter (dotted lines) of the gas density
(top row), of the gas temperature (middle row) and of the relative gas
velocity (bottom row) within the numerically resolved accretion region
versus BH mass at $z=2.8$ (green),  $z=1$ (orange) and $z=0$
(red) for the 500Mpc/hr run. {The predictions of the 68Mpc/uhr run
  are illustrated by the dashed lines in the left column.} All
quantities are in physical units. The mean relative gas velocity was
calculated by averaging over the relative velocities between the BH
and all the gas particles of $x,y$ and $z$-components and then by
computing the norm of the mean $x,y$ and $z$-velocity. 
To be consistent with our earlier results for the bolometric AGN
luminosities, we have only considered AGN with $\log(L_{\mathrm{bol}})
> 43$~[erg/s] (left column) and additionally distinguished between an
accretion in the quasar (middle column) and the radio mode (right
column), depending on the Eddington-ratio. 

Irrespectively of the redshift, both the mean gas temperature $T$ and
the mean relative velocity $V_{\mathrm{rel}}$ are increasing, while
the gas density $\rho$ is decreasing with increasing BH mass
(stronger at lower z) when considering \textit{all} AGN (left column)
in both simulation runs. This means -- when taking into account
Eq. \ref{Bondi} -- that at any redshift \textit{the``relative''
  accretion rates (i.e. accretion rate $/M_\bullet^2$)} tend to
decrease with increasing BH mass. This can explain the BH mass
dependence of the AGN number density at a given redshift as seen in
the AGN luminosity-black hole mass plane: AGN number densities of
massive BHs peak at low  Eddington-ratios and vice versa (see
Fig. \ref{Lum_bh}).   

Turning now to the redshift dependence of the average gas quantities
(coloured lines in the left column of Fig. \ref{Dens_bhmass}),    
we find for massive BHs ($M_\bullet > 10^7 M_\odot$) a
significant decrease of the mean gas density by almost three orders of
magnitudes from $z=2.8$ to $z=0$, which most likely causes a decrease
of the relative accretion rates at a given BH mass towards 
lower redshifts (even if the gas temperature, and thus the sound
speed, is also slightly decreasing towards lower redshifts). This can,
therefore, explain the decrease in the number density of very luminous
AGN with $\log(L_{\mathrm{bol}}) \geq 46$~[erg/s] (i.e. very massive
BHs $M_\bullet \geq 10^8 M_\odot$ accreting at high
Eddington-ratios) since $z \sim 2$ (see yellow and red curves in
Fig. \ref{Num_dens}). A further consequence of the decreasing
accretion rates of massive BHs with decreasing redshift is that these
objects have now preferentially moderate luminosities and thus, can
contribute to the weakly increasing number density of
\textit{moderately luminous} AGN with evolving time (see black and
dark/light blue lines in Fig. \ref{Num_dens}).

{Instead, for low mass BHs ($M_\bullet < 10^7 M_\odot$), the
  density, the gas temperature and the relative velocity (between the
  gas and the BH) are all slightly decreasing down to $z=1$ in the
  500Mpc/hr run. However, after $z=1$, the gas density is hardly
  decreasing anymore, the gas temperature stays constant, and the
  relative velocity is slightly decreasing suggesting that the
  accretion for low mass BHs is approximately constant or even
  slightly increasing with decreasing redshift. This ``not-decreasing''
  accretion in low-mass BHs with decreasing redshift (after $z=1$) can
  also contribute to the weak increase of the number densities of
  moderately luminous AGN (see black and dark blue lines in
  Fig. \ref{Num_dens}) -- in addition to the massive BHs accreting at
  low Eddington-ratios.}

{Considering the higher resolved 68Mpc/uhr run, the gas density
  in such low-mass BHs is higher, while the relative velocity is lower
  than in the 500Mpc/hr run. This leads to a stronger and faster growth of
low mass black holes, as e.g. seen for the low mass end of the BH mass
function. This is also responsible for the increase of the
low-luminous end of the AGN luminosity function in the 68Mpc/uhr
run. In addition, these trends suggest that the $\alpha$ parameter in
eq. \ref{Bondi} needs to be adjusted when going to higher and higher
resolution, as also discussed in \citet{Booth11} and
\citet{Vogelsberger13}.} 

Overall, the redshift and mass evolution of the different gaseous
quantities contributing to the Bondi-accretion can be a result of the
interplay of radio- and quasar-mode accretion as adopted in our model: 
At high redshifts $z>2$ for the whole mass range and at lower
redshifts for low mass black holes ($M_\bullet \leq  10^7 M_\odot$),
the gaseous quantities are driven by the evolution in the quasar mode
(see middle column of Fig. \ref{Dens_bhmass}). This is the dominating
accretion mechanism for these regimes and thus, is responsible for the
overall decrease of the mean densities, temperatures and relative
velocities of low mass BHs towards low redshifts. Instead, for low
redshifts $z \leq 1$ \textit{and} more massive BHs ($M_\bullet \geq
10^7 M_\odot$), the accretion in the radio-mode is dominating the
gaseous quantities (right column of Fig. \ref{Dens_bhmass}).

{The AGN feedback (in the radio-mode) together with the gas
  consumption due star formation in massive galaxies over cosmic time
  leads to a significantly decreasing gas density with increasing BH
  mass and decreasing redshift.} This is not the case for an accretion
at the quasar-mode, where at $z=0$ the gas density stays constant with
increasing BH mass and for the most massive BHs, the gas density tends 
to be even slightly larger at $z=0$ than at $z=1$. 

To summarise, we find that the downsizing trend in the AGN number  
density evolution and thus, the BH accretion, in our simulations is
mainly caused by ``global'' gas-physical properties in the vicinity of
the resolved accretion region and not necessarily by the ``local'' gas
properties within the (unresolved) Bondi accretion
radius. Interestingly, this is in agreement with detailed,
high-resolved (spatial resolution of 200~pc) 3D simulations of black
hole accretion, which indicate that the BH accretion is mainly driven
by large-scale gas properties beyond 1~kpc (G. Novak, personal
communication; Novak, Durier \& Babul, in prep.) and \textit{not by
  the local gas properties}. 

Essentially, we have demonstrated that a combination of both \textit{a
  decreasing mean gas density of  BHs with increasing BH mass and
  decreasing redshift} and \textit{a slightly lower mean relative
  velocity of low mass BHs at lower redshifts} (the latter being of
minor importance) is causing the downsizing trend in the AGN number
density evolution.

\section{Summary and discussion}\label{downdis}

In this study, we have analysed a subset of hydrodynamic, fully
cosmological simulations from the Magneticum Pathfinder simulation set      
(Dolag et al., in prep.) focusing on the evolution of statistical
properties of black hole growth and AGN luminosities from $z=5$ down
to $z=0$, in particular the observed downsizing trend in BH
growth. { The simulations are based on an improved SPH code
  Gadget3 (\citealp{Springel05gad}) where we are additionally using a
  higher order kernel based on the bias-corrected, sixth-order
  Wendland kernel  \citep{2012MNRAS.425.1068D} with 295 neighbours
  which together with a low-viscosity SPH scheme allows us to properly
  track turbulence within galaxy clusters
  \citep{2005MNRAS.364..753D,2013MNRAS.429.3564D}.}

The code also contains radiative gas cooling, a spatially uniform UV
background and a multi-phase model for star formation with the
associated feedback processes (\citealp{Springel03}). The simulations
also include a detailed model of chemical evolution according to
\citet{Tornatore07}, where metals are released by SNII, SNIa
explosions and AGB stars. {We also allow for isotropic thermal
  conduction with $1/20$ of the classical Spitzer value
  \citep{2004ApJ...606L..97D}.} In addition, the prescriptions for BH
growth and feedback from AGN are based on the models presented in
\citet{Springel05b} and \citet{DiMatteo05}, but contain the same 
modifications {for a transition from a quasar- to a radio-mode
  feedback as in the study of \citet{Fabjan10}. It also contains some
  new changes (as discussed in section \ref{BHgrowth}), where the most
  noticeable difference is reflected in our ability to follow BHs
  properly in galaxies which are inside of galaxy clusters.} 

We have considered two simulation runs, one with a large co-moving
volume of $(500\ \mathrm{Mpc})^3$ and one with a smaller volume of
$(68\ \mathrm{Mpc})^3$ but a higher resolution with initial
condistions based on the WMAP7 cosmology. We now summarise our main
results:

\begin{itemize}
\item[{\bf 1.}] Consistent with previous studies
  (e.g. \citealp{DiMatteo08}), our simulations are in reasonably good
  agreement with the observed present-day BH-stellar mass relation and
  the BH mass function, although the high-mass end is over-estimated
  due to too inefficient radio mode feedback. The latter point is also
  true for the stellar mass function, i.e. the simulations tend to
  produce too many too massive galaxies hosting too massive BHs after
  $z=1$. The main growth of BHs is found to occur before $z \sim 1$
  when considering the evolution of the BH mass function, which is
  largely in place at $z\sim 1$ in qualitative agreement with
  observational estimates for the evolution of the BH mass function
  (\citealp{Merloni08}).  This is also consistent with the cosmic
  evolution of BH accretion rate densities, which peak around $z \sim
  1.5 $ and start to strongly decline towards lower redshifts.     

\item[{\bf 2.}] {Our large-volume simulation can successfully
    reproduce the observed bolometric AGN luminosity function up to $z
    \sim 3$ (\citealp{Hopkins07}), even if moderately luminous AGN are
    slightly over-estimated at $z=1.5=2.5$. For moderately luminous
    AGN this is consistent with a previous study of \citet{Degraf10},
    \textit{but the high luminosity end ($\log(L_{\mathrm{bol}}/L_\odot) \geq
      45$~erg/s) matches -- for the first time -- the observational
      data}. At higher redshifts (up to $z \sim 5$), however, the
    large-volume simulation predicts a too small amount of AGN due to
    resolution effects. Turning to the run with a smaller volume but
    higher resolution helps to increase the amount of moderately
    luminous AGN in reasonably good agreement with observations. Even
    if the smaller volume, higher resolution simulation predicts more
    moderately luminous AGN than the large volume run, it {converges
    against} the observational data for more luminous AGN  ($\log
  (L_{\mathrm{bol}} /L_\odot)>45$~erg/s). }  

\item[{\bf 3.}] Besides the bolometric luminosities, we have
  additionally extended our simulation predictions by dust obscuration
  (in post-processing) to compare them directly with recent soft and
  hard X-ray measurements of the AGN luminosity function. When
  adopting an empirically motivated dust obscuration model, where
  obscuration of AGN is dependent on both, redshift and luminosity, we
  find a reasonably good agreement between the large-volume simulation
  and observations up to $z \sim 3$ and up to a redshift $z=4-5$
  for the small-volume, but higher resolution simulation.  

  Dividing our AGN in low and high excitation state, we are also
  able to reproduce observed number densities of present day
  LERGs/HERGs radio galaxies, implying that the simulations can
  successfully capture the correct amount of radiatively efficient and
  inefficient AGNs. 

\item[{\bf 4.}] As a consequence of points 2. and 3., a combination of
  both simulation runs (small- and large-volume) can successfully
  reproduce the co-moving number density evolution of AGN up to
  $z=4-5$,  and therefore, naturally predict the characteristic,
  observed downsizing trend within a hierarchical structure formation
  scenario. The strong decline of luminous AGN with decreasing
  redshift (since $z=2$) can be attributed to the gas density in the
  vicinity of a (massive) BH, which gets successively depleted
  with evolving time due star formation and AGN feedback. The gas
  inflow towards a massive BH is inhibited and its luminosity starts
  to fade (and can then contribute to the amount of moderately
  luminous AGN). This also explains why massive BHs preferentially
  accrete at low Eddington-ratios after $z = 2$. 

  The continuous increase of moderately luminous AGN over time can,
  instead, be seen a result of both massive BHs in their ``fading''
  phase and lower mass BHs as they tend to have constant or slightly
  higher accretion rates towards lower redshifts (after $z=1$) due to
  slightly decreasing mean relative velocities between the BH and the
  surrounding gas with decreasing redshift. 
 
\item [{\bf 5.}] Finally, we find that the downsizing behaviour implies
  that the peaks of the Eddington ratio distributions are shifted
  towards successively smaller Eddington ratios with decreasing
  redshift -- a trend which has also been found by observational
  studies. The number of BHs accreting close to the
  Eddington-rate is more than one order of magnitude smaller at
  $z=0$ than at high redshifts ($z=1-3$), where the BHs are
  preferentially accreting within a broad range of Eddington ratios
  $0.01<f_{\mathrm{edd}}<1$. Besides, our simulations can successfully
  predict the observed evolution of the mean Eddington-ratio for AGN
  more luminous than $L_{\mathrm{bol}} > 10^{43}$~[erg/s]. These
  implications of the downsizing behaviour are in qualitative
  agreement with predictions from semi-analytic models (see
  e.g. \citealp{Fanidakis12, Hirschmann12}).
\end{itemize}

Despite the overall success of the simulations being in reasonably
good agreement with observational data we should keep in mind that
this is attained despite of the rather crude, approximate and
numerically limited description of BH accretion, energy
extraction and thermalisation (even if high-resolution simulations of
BH accretion indicate an independence of the BH
accretion on local gas properties up to 1~kpc around the BH;
Novak, Durier and Babul, in prep.). In addition, the simulations
presented in this work fail to reproduce the massive end of the BH and
the stellar mass functions at low redshift due to too inefficient
radio-mode feedback.

But nevertheless, this study nicely demonstrates that the observed
downsizing trend in BH growth is not contradictory with a hierarchical
structure formation model (consistent with the predictions of
semi-analytic models as shown in \citealp{Bonoli09, Fanidakis12,
  Hirschmann12}), but instead is a ``natural'' outcome due to specific
baryon processes.  Besides the ``downsizing'' trend, there exist many
other unresolved questions about BH growth and AGN evolution which can
be easily assessed in our simulations. Therefore, in forthcoming
studies, we plan to examine the connection between AGN and their host
galaxy properties, AGN clustering properties and typical AGN light
curves (of differently luminous AGN) which may particularly
illuminate our current understanding of how strongly galaxies and BHs
are co-evolving and which are the main trigger mechanisms for AGN
activity (if there are any). Particularly the latter is -- up to now
-- a major unresolved open issue in understanding BH growth and AGN 
evolution. 

\section*{Acknowledgments}
We thank Andrea Merloni, Francesco Shankar, James Aird and Jacobo
Ebrero for providing us with observational data and Andrea 
Merloni, Greg Novak and Rachel Somerville for fruitful discussions.

This research was supported by the DFG Cluster of Excellence 'Origin
and structure of the universe'. {We are especially grateful for the 
support by M. Petkova through the Computational Center for Particle
and Astrophysics (C$^2$PAP). Computations have been performed at the
at the 'Leibniz-Rechenzentrum' with CPU time assigned to the
Project 'pr86re' as well as at the 'Rechenzentrum der Max-Planck-
Gesellschaft' at the 'Max-Planck-Institut f\"ur Plasmaphysik' with
CPU time assigned to the 'Max-Planck-Institut f\"ur Astrophysik'. }
M.H. acknowledges financial support from
the European Research Council under the European Community's Seventh
Framework Programme (FP7/2007-2013)/ERC grant agreement
n. 202781. S.B. acknowledges financial support from the European
Commission's Framework Programme 7, through the Marie Curie Initial
Training Network Cosmo-Comp (PITN-GA-2009-23856), the PRIN-MIUR-2009   
grant ``Tracing the growth of structures in the Universe'' and the
PD51-INFN grant.

\bibliographystyle{mn2e}
\bibliography{Literaturdatenbank}

\label{lastpage}

\end{document}